\documentclass{article}

\PassOptionsToPackage{numbers,compress}{natbib}
\usepackage[preprint]{neurips_2026}

\usepackage[utf8]{inputenc}
\usepackage[T1]{fontenc}
\usepackage{hyperref}
\usepackage{url}
\usepackage{booktabs}
\usepackage{amsfonts}
\usepackage{amsmath}
\usepackage{nicefrac}
\usepackage{microtype}
\usepackage{xcolor}
\usepackage{colortbl}
\usepackage{graphicx}
\usepackage{subcaption}
\usepackage{multirow}
\usepackage[inline]{enumitem}
\usepackage{placeins}
\usepackage{tikz}
\usetikzlibrary{positioning, arrows.meta, calc}
\usepackage{fontawesome5}
\title{TuneJury: An Open Metric for Improving\\
Music Generation Preference Alignment}

\author{%
  Yonghyun Kim$^{\sharp}$ \\
  \And
  Junwon Lee$^{\flat\flat}$ \\
  \And
  Haiwen Xia$^{\natural\natural}$ \\
  \And
  Yinghao Ma$^{\sharp\sharp}$ \\
  \AND
  Junghyun Koo$^{\natural}$ \\
  \And
  Koichi Saito$^{\natural}$ \\
  \And
  Yuki Mitsufuji$^{\natural}$ \\
  \And
  Chris Donahue$^{\flat}$ \\
  \And
  \vspace{0.5em}
  \small
  \begin{tabular}{@{}c@{}}
    $^{\flat}$Carnegie Mellon University \quad
    $^{\natural}$Sony AI \quad
    $^{\sharp}$Georgia Tech \\
    $^{\flat\flat}$KAIST \quad
    $^{\natural\natural}$Peking University \quad
    $^{\sharp\sharp}$QMUL
  \end{tabular}
}

\begin{document}

\maketitle

\begin{abstract}

We introduce \textbf{TuneJury}, an open, instance-level pairwise reward model for text-to-music that predicts a music preference score from a text prompt and an audio clip. The released checkpoint is trained on publicly available human-preference labels covering arena-style (A~vs.~B) votes, metric-alignment preference pairs, crowdsourced pairwise comparisons, and expert aesthetic ratings. The predicted score margin between two clips is well calibrated on our held-out test split, supporting data filtering via a simple score threshold. TuneJury generalizes to both held-out test pairs and out-of-distribution benchmarks, remaining competitive with prior baselines on the latter. For generators released after training, we introduce \emph{anchor calibration}, a post-hoc, per-system Bradley--Terry calibration that recovers agreement at substantially better data efficiency than from-scratch retraining. The same frozen reward drives consistent reward-axis gains across three downstream applications: inference-time best-of-$N$ selection, DITTO-style latent optimization, and expert-iteration post-training.
\end{abstract}

\vspace{-4mm}
\begin{center}
\textbf{TuneJury is available at \url{https://github.com/yonghyunk1m/TuneJury}.}
\end{center}

\section{Introduction}
\label{sec:intro}

Music preference is subjective~\cite{cideron2024musicrl}, which makes text-to-music (TTM) evaluation difficult. Popular metrics such as Fr\'echet audio distance (FAD)~\cite{kilgour2019fad} and its encoder-specific variants~\cite{gui2024fad} do not address this directly: they measure encoder-space similarity to a reference set rather than human preference, and they describe collections rather than individual clips~\cite{huang2025musicprefs}. Even the same TTM system produces variable quality from one generation to the next. To choose which generation a listener prefers, to track how the model performs across prompts and genres, or to pick which samples to fine-tune on next, we need a \emph{per-clip} evaluation metric that reflects human preference. Absolute mean opinion score (MOS) regression, an alternative used in adjacent audio domains such as speech~\cite{lo2019mosnet}, does score clips one at a time. However, its assumption that raters share a scale is fragile for any subjective rating task: scales drift across sessions and individuals~\cite{rosenberg2017bias, cooper2023mos}. This drift is especially strong for music, where preference depends on individual taste. An absolute regressor therefore inherits this drift as systematic noise. Pairwise A~vs.~B comparison avoids the shared-scale assumption altogether: each rater stays within their own scale, yielding lower measurement variance than direct rating~\cite{mantiuk2012comparison}. Pairwise modeling captures preference as a population-level probability rather than an absolute quality, which is how a subjective signal admits a well-defined score.

A model trained on such comparisons to predict human preference is a \emph{reward model}. The design was introduced in deep reinforcement learning~\cite{christiano2017deeprl} and is now standard in language model alignment, where it supplies the training signal for reinforcement learning from human feedback~\cite{ouyang2022instructgpt}. The paradigm has recently reached speech naturalness judgment (SpeechJudge~\cite{zhang2026speechjudge}). In music, the most directly comparable prior work is CMI-RewardModel (CMI-RM), from CMI-RewardBench~\cite{ma2026cmirewardbench}. To our knowledge, CMI-RM is the only existing music reward model trained with the shared-weight pairwise-logistic setup of RankNet~\cite{burges2005ranknet}. It consumes text, lyrics, reference audio, and candidate audio, with a $2$-axis (alignment, quality) output, trained on ${\sim}110$\,K LLM pseudo-labels augmented with ${\sim}6.6$\,K human-provided pairs. This raises two natural questions. First, \emph{how well can a leaner reward model that scores only (prompt, audio) data points, without lyrics or reference audio, perform on the same task?} Second, \emph{can a model trained on human pairs without pseudo-label augmentation reach competitive accuracy?} 

To probe both, we introduce \textbf{TuneJury}, a small MLP head over frozen audio and text encoders that scores a (prompt, audio) data point with a single music preference scalar, at ${\sim}2.8$\,M trainable parameters vs.\ CMI-RM's ${\sim}30$\,M. We train TuneJury on ${\sim}17.5$\,K training pairs from four open human-rated sources (Music Arena~\cite{kim2025musicarena}, MusicPrefs~\cite{huang2025musicprefs}, AIME~\cite{grotschla2025aime}, SongEval~\cite{yao2025songeval}; Section~\ref{sec:tunejury}). On the CMI-RewardBench Music Arena split, TuneJury's pairwise accuracy is on par with the two authors' agreement with the released vote on a $30$-pair human-ceiling probe of the same split (Section~\ref{sec:eval_internal}). TuneJury also substantially outperforms the no-pseudo CMI-RM ablation (CMI-RM trained on its ${\sim}6.6$\,K human pairs alone, without the ${\sim}110$\,K pseudo-labels) on PAM~\cite{deshmukh2024pam} and MusicEval~\cite{liu2025musiceval} Spearman rank correlation coefficient (SRCC), and stays competitive with the full pseudo-augmented CMI-RM on out-of-distribution (OOD) splits. This leaner setup is deliberate: CMI-RM's lyrics and reference-audio inputs go unused in our instrumental-music scope, so its effective input collapses to text $+$ audio like TuneJury, while we leave pseudo-label augmentation to follow-up work.

\begin{table}[!ht]
\vspace{-0.5em}
\centering
\footnotesize
\setlength{\tabcolsep}{4pt}
\caption{Design comparison of the six music reward~/~quality scorers evaluated in this paper (Section~\ref{sec:eval_external}), all built on frozen pretrained backbones. TuneJury and CMI-RM share the RankNet pairwise paradigm but differ in input scope, output structure, and supervision. Input codes (matching Table~\ref{tab:head_to_head}): T~$=$~text prompt, L~$=$~lyrics, R~$=$~reference audio (optional style or continuation input, used by CMI-RM), A~$=$~candidate audio.}
\label{tab:design_compare}
\begin{tabular}{lllll}
\toprule
& Framework & Input  & Output & Supervision \\
\midrule
\textbf{TuneJury (ours)} & RankNet pairwise & TA & 1-d scalar & ${\sim}17.5$\,K human-rated pairs \\
CMI-RM~\cite{ma2026cmirewardbench} & RankNet pairwise & TLRA & 2-d (align, qual)  & ${\sim}6.6$\,K human $+$ ${\sim}110$\,K pseudo \\
SongEval-RM~\cite{yao2025songeval} & MOS regression & A    & 5-d aesthetic & SongEval MOS \\
Audiobox-Aesthetics~\cite{tjandra2025audiobox} & MOS regression & A & 4-d aesthetic & Audiobox MOS \\
MuQ-Eval~\cite{zhu2026muqeval} & MOS regression & A & 2-d (align, qual) & MusicEval MOS \\
PAM score~\cite{deshmukh2024pam} & Zero-shot audio-LM & TA     & 1-d scalar & zero-shot \\
\bottomrule
\end{tabular}
\end{table}

Beyond benchmark accuracy, we exercise TuneJury as a \emph{preference-alignment signal}: the same frozen reward drives consistent reward-axis gains across three downstream applications (Section~\ref{sec:applications}), using no additional human labels. \emph{(i) Mode~1: inference-time best-of-$N$ selection.} On four frozen open-weights backbones, Top-$1$ reward stays strictly monotone in $N$ through $N{=}32$ (Appendix~\ref{app:mode1_extended}). \emph{(ii) Mode~2: DITTO-style latent optimization.} DITTO-style~\cite{novack2024ditto} optimization lifts mean reward on both SAO-small~\cite{novack2025saoarc} and TangoFlux~\cite{hung2025tangoflux}. The low-reward TangoFlux baseline moves closer to a music reference set and improves text-audio alignment, while the higher-reward SAO-small baseline drifts away on both distributional and alignment side metrics, exposing the classic reward-exploitation pattern~\cite{gao2023scaling}, in which gains on a learned reward come at the expense of other quality measures. \emph{(iii) Mode~3: expert-iteration post-training.} Expert iteration~\cite{anthony2017expertiteration, singh2024rest} on a rectified-flow DiT~\cite{peebles2023dit} traces a reward-fidelity Pareto trade-off across three fine-tune learning rates, exposing the same pattern.

After a reward model is trained, new TTM systems keep being released. Music from different systems carries different characteristics (timbre, mixing, style choices), so a reward model trained on existing systems may score music from a new system uniformly above or below its trained scale. Realigning the new system's scores traditionally requires retraining with fresh human ratings, which is expensive. We additionally introduce \emph{anchor calibration}, a post-hoc, per-system Bradley--Terry calibration that matches retraining's accuracy ceiling on post-cutoff Music Arena battles with ${\sim}25\times$ less calibration data, allowing TuneJury to adapt to each new TTM system without retraining (Appendix~\ref{app:anchor_calibration}). All artifacts (checkpoints, code, demos) are openly released (Appendix~\ref{app:release_detail}).

\break
\paragraph{Contributions.}
\begin{itemize}[leftmargin=*,itemsep=2pt]
\item \textbf{TuneJury}: an instance-level music preference reward model trained pairwise on ${\sim}17.5$\,K human-rated A~vs.~B pairs from four open sources without pseudo-label augmentation. The released $2.8$\,M-parameter instance reaches $0.7086$ pairwise accuracy on a $2{,}035$-pair held-out test split, outperforms the no-pseudo CMI-RM ablation by ${+}0.17$ SRCC on PAM and MusicEval, and stays within $2$ percentage points (pp) of the full pseudo-augmented CMI-RM on OOD splits.
\item \textbf{Three downstream applications on a single frozen reward}: best-of-$N$ selection (Mode~1), DITTO-style latent optimization (Mode~2), and expert-iteration post-training (Mode~3). Across these, TuneJury delivers consistent reward-axis gains. The Mode~3 learning-rate sweep maps a tunable Pareto trade-off between reward gain and distributional fidelity.
\item \textbf{Anchor calibration}: a post-hoc, per-system Bradley--Terry calibration that matches retraining's accuracy ceiling on post-cutoff Music Arena battles with ${\sim}25\times$ less calibration data, allowing TuneJury to adapt to each new TTM system without retraining.
\item \textbf{Open release} of checkpoints, code, listening demos, and pre-computed reward scores on seven open-license music collections.
\end{itemize}
   
\section{Related Work}
\label{sec:related}

\paragraph{Music reward models.}
RankNet~\cite{burges2005ranknet} introduced the shared-weight pairwise-logistic learning-to-rank formulation, widely adopted by subsequent text-to-image preference reward models~\cite{xu2023imagereward, kirstain2023pickapic, wu2023hpsv2}. Within music reward modeling, the only prior work we are aware of that adopts the same setup is CMI-RM~\cite{ma2026cmirewardbench}, which passes both candidates of a pair through a shared backbone over text, lyrics, reference audio, and candidate audio. The four other music reward models we benchmark against use non-pairwise objectives: multi-axis MOS regression (Audiobox-Aesthetics~\cite{tjandra2025audiobox}, SongEval-RM~\cite{yao2025songeval}, and MuQ-Eval~\cite{zhu2026muqeval}) and zero-shot prompting of an audio-language model (PAM score~\cite{deshmukh2024pam}). Table~\ref{tab:design_compare} summarizes how TuneJury sits along four axes (framework, input scope, output structure, supervision). Head-to-head numbers against all five baselines on the CMI-RewardBench test splits appear in Section~\ref{sec:eval_external}. TuneJury shares the pairwise paradigm with CMI-RM, drops CMI-RM's lyrics and reference-audio channels, outputs a single preference scalar instead of a multi-axis vector or alignment-quality pair, and pools four open human-rated sources without pseudo-label augmentation.

\paragraph{Open music preference data.}
The four open human-labeled sources we pool, all newly released in 2025, are Music Arena~\cite{kim2025musicarena} (live arena pairwise battles), MusicPrefs~\cite{huang2025musicprefs} (pairwise preferences across fidelity and musicality axes), AIME~\cite{grotschla2025aime} (crowdsourced pairwise comparisons), and SongEval~\cite{yao2025songeval} ($5$-axis aesthetic ratings by professional musicians). CMI-RewardBench~\cite{ma2026cmirewardbench} is a benchmark for music reward models, accompanied by CMI-RM. Our Music Arena training pool overlaps CMI-RewardBench's Music Arena
test split, an overlap we remove from our pool before training the released checkpoint
(Section~\ref{sec:eval_external}, Appendix~\ref{app:cmi_external_detail}).

\paragraph{Other automated metrics for text-to-music generation.}
Current metrics fall mainly into three groups~\cite{ma2024foundation, lerch2025survey}. (i)~\emph{Distributional similarity}, dominated by FAD~\cite{kilgour2019fad} (a Fr\'echet distance in audio embedding space) and its encoder-specific variants (FAD-CLAP, FAD-MERT, etc.~\cite{gui2024fad}), typically paired with KL divergence on audio-classifier logits in MusicGen~\cite{copet2023musicgen} / AudioLDM2~\cite{liu2024audioldm2} / Stable Audio Open~\cite{evans2024stableaudio} evaluation, with Kernel Audio Distance (KAD)~\cite{chung2025kad} and MAUVE Audio Divergence (MAD)~\cite{huang2025musicprefs} as recent alternatives. (ii)~\emph{Text-audio alignment}, relying on the CLAP score~\cite{wu2023clap}. (iii)~\emph{No-reference quality prediction}, such as Audiobox-Aesthetics~\cite{tjandra2025audiobox} and PAM score~\cite{deshmukh2024pam}.

\paragraph{Preference learning and reward-driven post-training.}
Direct preference optimization (DPO)~\cite{rafailov2023dpo} and DPO-style audio counterparts (e.g., Tango~2~\cite{majumder2024tango2}, TangoFlux~\cite{hung2025tangoflux}) align generative models against external preference labels. Reward-driven fine-tuning for diffusion models splits into policy-gradient methods (estimating updates from sampled rewards without differentiating the sampler), including denoising diffusion policy optimization (DDPO)~\cite{black2024ddpo} and DPOK~\cite{fan2023dpok}, and reward-backprop methods (differentiating the reward through the sampling chain), including DRaFT~\cite{clark2024draft} and ReFL~\cite{xu2023imagereward}. Group-relative policy optimization (GRPO)~\cite{shao2024deepseekmath}, introduced for language model reasoning, has more recently been applied as another policy-gradient option in the diffusion setting~\cite{xue2025dancegrpo}. A separate line of work uses only the model's own samples and a frozen reward signal, spanning inference-time selection, latent optimization, and own-sample fine-tuning: best-of-$N$ selection (Mode~1, standard in text-to-image~\cite{xu2023imagereward} and language~\cite{gao2023scaling}), DITTO~\cite{novack2024ditto} inference-time latent optimization (Mode~2, backprop through the sampler into the noise latents), and expert iteration~\cite{anthony2017expertiteration} (Mode~3), whose LLM-domain variants include reward-ranked self-training (ReST)~\cite{gulcehre2023rest, singh2024rest}. Modes~1--3 (Section~\ref{sec:applications}) use the frozen TuneJury reward as their only supervision signal. No additional human preference labels are collected.

\section{TuneJury}
\label{sec:tunejury}

TuneJury is an instance-level pairwise reward model for text-to-music. A small MLP head reads frozen audio and text embeddings and maps a single (prompt, audio) data point to a scalar preference score, trained with the shared-weight pairwise-logistic objective on human A~vs.~B judgments and competitive on CMI-RewardBench (Section~\ref{sec:evaluation}). We describe the inputs and head architecture, the frozen encoder stack and its robustness to the encoder choice, the four open human-rated training sources, and the pairwise training procedure.

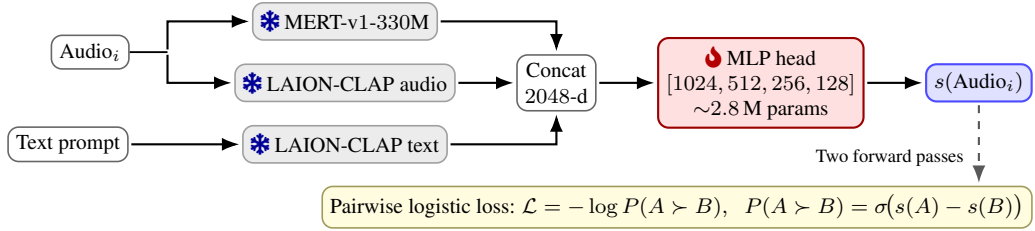
\begin{figure}[!t]
\centering
\tikzset{
    base/.style={draw, rounded corners, font=\footnotesize, align=center, inner sep=3pt, line width=0.6pt},
    inp/.style={base, fill=white, draw=black!60},
    frozen/.style={base, fill=gray!15, draw=gray!75},
    cat/.style={base, fill=white, draw=black!60},
    trainable/.style={base, fill=red!12, draw=red!60!black, line width=0.8pt},
    output/.style={base, fill=blue!12, draw=blue!70, line width=0.8pt},
    lossbox/.style={base, fill=yellow!15, draw=yellow!55!black},
    arr/.style={->, >=Latex, thick, shorten >=1pt, shorten <=1pt},
}
\providecommand{\snow}{{\color{blue!60!black}\faSnowflake}}
\providecommand{\fire}{{\color{red!70!black}\faFire}}
\begin{tikzpicture}

% Three frozen encoders (anchor spine)
\node[frozen] (mert) {\snow~MERT-v$1$-$330$M};
\node[frozen, below=3mm of mert] (clap_a) {\snow~LAION-CLAP audio};
\node[frozen, below=3mm of clap_a] (clap_t) {\snow~LAION-CLAP text};

% Inputs: audio between mert/clap_a, text aligned with clap_t
\coordinate (audio_y) at ($(mert.west)!0.5!(clap_a.west)$);
\node[inp, anchor=east] (audio) at ($(audio_y) + (-15mm, 0)$) {Audio$_i$};
\node[inp, anchor=east] (text) at ($(clap_t.west) + (-15mm, 0)$) {Text prompt};

% Y-junction for audio: short stub then branch up/down
\coordinate (Y) at ($(audio.east) + (5mm, 0)$);
\draw[line width=0.7pt] (audio.east) -- (Y);
\draw[arr] (Y) |- (mert.west);
\draw[arr] (Y) |- (clap_a.west);
\draw[arr] (text.east) -- (clap_t.west);

% Right pipeline (wider spacing to fill linewidth)
\node[cat, right=8mm of clap_a] (concat) {Concat\\$2048$-d};
\node[trainable, right=8mm of concat] (mlp) {\fire~MLP head\\$[1024, 512, 256, 128]$\\${\sim}2.8$\,M params};
\node[output, right=8mm of mlp] (s) {$s(\text{Audio}_i)$};

% Encoders -> Concat
\draw[arr] (mert.east) -| (concat.north);
\draw[arr] (clap_a) -- (concat);
\draw[arr] (clap_t.east) -| (concat.south);

% Pipeline
\draw[arr] (concat) -- (mlp);
\draw[arr] (mlp) -- (s);

\node[lossbox, anchor=north east] (loss) at ($(s.east |- s.south) + (0, -11mm)$)
    {Pairwise logistic loss: $\mathcal{L} = -\log P(A \succ B), \;\; P(A \succ B) = \sigma\!\big(s(A) - s(B)\big)$};

\draw[arr, dashed, gray!70!black] (s.south) --
    node[pos=0.66, left, font=\scriptsize, black]{Two forward passes\;}
    (s.south |- loss.north);

\end{tikzpicture}
\caption{TuneJury architecture (CLAP$+$MERT instantiation). Audio$_i$ feeds two audio encoders (MERT-v$1$-$330$M, LAION-CLAP audio) and the text prompt feeds a text encoder (LAION-CLAP text), with all three frozen (\snow). The $2048$-d concatenated embedding passes through a trainable (\fire) MLP head with ${\sim}2.8$\,M parameters, producing scalar $s(\cdot)$. The head has shared weights across both clips ($A$, $B$), trained with the pairwise logistic loss. Inference scores one clip per pass.}
\label{fig:tunejury_arch}
\end{figure}

\paragraph{Inputs.}
In the CLAP$+$MERT instantiation shown in Figure~\ref{fig:tunejury_arch}, the $2048$-d input concatenates three pre-extracted embeddings chosen for complementary roles.

\FloatBarrier

\paragraph{Encoder details.}
LAION-CLAP-Music's $512$-d audio and $512$-d text embeddings (\href{https://huggingface.co/lukewys/laion_clap/blob/main/music_audioset_epoch_15_esc_90.14.pt}{\texttt{music\_audioset\_epoch\_15\_esc\_90.14}} checkpoint~\cite{wu2023clap}; LAION-CLAP for short below) provide paired text/audio features. The $1024$-d \href{https://huggingface.co/m-a-p/MERT-v1-330M}{MERT-v$1$-$330$M} audio embedding~\cite{li2024mert} provides a music-pretrained audio representation. The three are concatenated in the order $[\text{CLAP audio}, \text{MERT audio}, \text{CLAP text}]$ (Figure~\ref{fig:tunejury_arch}). The input ablation in Appendix~\ref{app:feature_modality} reports each embedding's standalone and combined contribution. For any clip without a prompt, including SongEval training pairs and empty-prompt inference, the text branch receives a $512$-d zero vector in place of the CLAP text embedding. LAION-CLAP audio and text vectors use the model's default pooling (variable-length input collapsed to $512$-d). The MERT vector is the time-mean of its final hidden state.

\paragraph{Training objective and architecture.}
The scoring head is a small $4$-hidden-layer MLP (widths $[1024, 512, 256, 128]$, ${\sim}2.8$\,M trainable parameters) over the $2048$-d concatenated input. It follows the shared-weight pairwise-logistic setup introduced by RankNet~\cite{burges2005ranknet} (also adopted by CMI-RM~\cite{ma2026cmirewardbench}; Section~\ref{sec:related}):
the head outputs a raw score $s(\cdot)$, the win probability between paired clips is $P(A \succ B) = \sigma(s(A) - s(B))$, and we minimize the binary cross-entropy against the preference label (ties take a soft label of $0.5$). Absolute scores are not anchored to a fixed scale (the pairwise-logistic loss is shift-invariant), and margins between distinguishable pairs can grow without bound. Training-mix scores mostly fall within $[-2,+2]$, with the largest pair margins reaching ${\sim}10$ on SongEval high-margin synthesized pairs (Appendix~\ref{app:calibration_bins}). The per-dataset distribution on the released human-music collections is narrower (the $10$th to $90$th percentile range is contained within $[-2.6, +2.4]$ across the seven sources; Figure~\ref{fig:release_distribution}).

\paragraph{Encoder robustness.}
The MLP head template is robust to the choice of music-pretrained encoder. Holding the head template (hidden widths scaled to the encoder dimension, halved to $[512, 256, 128, 64]$ for the $1024$-d MuQ-MuLan input), training procedure, and training mix (Music Arena excluded, three datasets retained) fixed, swapping CLAP$+$MERT for \href{https://huggingface.co/OpenMuQ/MuQ-MuLan-large}{MuQ-MuLan-large}'s $1024$-d joint embeddings~\cite{tencent2025muq} matches or beats the CLAP$+$MERT baseline on four of five OOD axes (Appendix~\ref{app:cmi_external_detail}, ``Encoder swap probe''). The CLAP$+$MERT instantiation is the reward signal across every Mode~1--3 demonstration in Section~\ref{sec:applications}.

\paragraph{Training data.}
We pool human-labeled data from four open sources (Table~\ref{tab:training_data}). AIME additionally includes MTG-Jamendo~\cite{bogdanov2019mtg} as a real-music baseline ($2{,}400$ of $15{,}600$ pairs). The first three sources are released as pairwise comparisons. SongEval contains instance-level annotations: we synthesize pairs via a $\ge 0.5$ mean-gap filter on its $5$ aesthetic axes ($3{,}760$ pairs), then assign songs to train/val/test and drop cross-split pairs, leaving $2{,}986$ pairs ($2{,}491$ train, $246$ val, $249$ test).

\begin{table}[!ht]
\centering
\footnotesize
\setlength{\tabcolsep}{4pt}
\caption{TuneJury training data sources. \emph{Pairs}: post-filter count used in our train/val/test splits. \emph{Prompt}: whether each pair has a text prompt.}
\label{tab:training_data}
\begin{tabular}{cccccc}
\toprule
Dataset & Pairs & Label type & \# TTM Systems & Raters & Prompt \\
\midrule
Music Arena~\cite{kim2025musicarena}    & $699$    & Live A~vs.~B battles         & $14$                 & Music Arena users    & Yes \\
MusicPrefs~\cite{huang2025musicprefs}   & $2{,}515$  & Metric-alignment A~vs.~B     & $7$                  & Crowdworkers           & Yes \\
AIME~\cite{grotschla2025aime}           & $15{,}600$ & Crowdsourced A~vs.~B         & $12$                  & Crowdworkers           & Yes \\
SongEval~\cite{yao2025songeval}         & $2{,}986$  & $5$-axis MOS $\to$ pairs & $5$                  & Professional musicians & No  \\
\bottomrule
\end{tabular}
\end{table}

\paragraph{Optional text input.}
Text prompts are optional for TuneJury: the metric can produce a score for an audio input alone. We use this capability for SongEval, which releases audio and aesthetic ratings but not the prompts used to generate the audio: its text branch receives a $512$-d zero vector during training. The other three sources release prompts, which we feed through the CLAP text encoder. One nuance: MusicPrefs annotators rated pairs without seeing the prompts~\cite{huang2025musicprefs}, so the released ratings are not text-conditioned. For a uniform input pipeline, we still pass MusicPrefs prompts through the text branch; an ablation that zero-vectors them instead tracks the released variant within ${\pm}3$\,pp on every external axis (PAM, MusicEval, CMI-Pref, Music Arena) with mixed signs.

This naturally splits the score into an audio-only part (musicality) and the text branch's contribution (text alignment). We probe this decomposition on CMI-RewardBench's PAM and MusicEval per-axis MOS pool, separate from TuneJury's ${\sim}17.5$\,K-pair preference training (four-stage protocol in Appendix~\ref{app:decomp_probe}). Subtracting the audio-only score from the text+audio score does not recover alignment. A small fresh head trained on the ${\sim}900$-clip alignment-labeled pool, however, reaches SRCC $0.444$ on the held-out component of alignment MOS not linearly explained by musicality ($20$-seed mean, $95\%$ confidence interval (CI) above zero), and the data-scaling curve has not yet plateaued at the pool's size limit. The probe makes no claim about TuneJury's main training data scale. Scaling strategies are discussed in Section~\ref{sec:discussion} (Open directions).

\paragraph{Bench-clean Music Arena.}
Clip-level labels are used end-to-end without chunking. For Music Arena, our pool spans battles dated 2025-07 to 2026-01 after dropping battles with missing audio outputs or \texttt{BOTH\_BAD} vote outcomes (\texttt{TIE} outcomes are retained with soft label $0.5$). From this pool (train, validation, and held-out test), we further remove every \texttt{battle\_uuid} that appears in CMI-RewardBench's $1{,}340$-pair Music Arena test split, so both our training and our held-out test are item-level disjoint from the CMI-RewardBench Music~Arena split ($131$ pairs removed from our internal Music Arena test, leaving $74$ pairs total, of which $20$ are non-tie and used for binary accuracy). We refer to this overlap-free pool and the resulting checkpoint as ``bench-clean'' throughout the paper. Distributional shift over time (newer generators entering after our training cutoff) is probed separately in Appendix~\ref{app:anchor_calibration}. After bench-overlap removal and per-dataset splits, the mix has ${\sim}22$\,K total pairs ($17{,}554$ training, $2{,}111$ validation, $2{,}135$ held-out test). Binary accuracy and expected calibration error (ECE)~\cite{guo2017calibration} use the $n{=}2{,}035$ non-tie subset.

\paragraph{Training procedure.}
We train the MLP head only (encoder features pre-extracted) with the AdamW optimizer~\cite{loshchilov2019adamw} and early stopping on validation loss. A full run completes in minutes on a single mid-range GPU. Full hyperparameters in Appendix~\ref{app:reproducibility}.

\section{Evaluation}
\label{sec:evaluation}

We report two evaluation settings. Section~\ref{sec:eval_internal} covers internal evaluation on the four-dataset held-out test split: pairwise accuracy, calibration, sanity checks on edge inputs, and input ablation. Section~\ref{sec:eval_external} benchmarks TuneJury against five prior reward models on CMI-RewardBench~\cite{ma2026cmirewardbench} splits disjoint from training: CMI-RM (the most direct pairwise comparison), three MOS regressors (Audiobox-Aesthetics, SongEval-RM, MuQ-Eval), and a zero-shot audio-language model (PAM score).

\subsection{Internal evaluation}
\label{sec:eval_internal}

\paragraph{Pairwise accuracy and calibration.}
On the $2{,}035$-pair held-out test split aggregated across our four training datasets (Section~\ref{sec:tunejury}, ties excluded), TuneJury reaches $0.7086$ pairwise accuracy\footnote{Human ceiling probe: two authors independently blind-labeled the same $30$-pair subset of the CMI-RewardBench Music Arena test split (item-disjoint from our training pool). Agreement with the released vote was $19/30$ ($0.633$) and $21/30$ ($0.700$), and inter-author agreement was $26/30$ ($0.867$). The $n{=}30$ sample is too small to establish a precise ceiling.} with ECE $0.0339$.\footnote{Weighted mean absolute gap between observed win rate and mean predicted confidence within each bin, over $10$ equal-count bins of $|s(A){-}s(B)|$. Reliability diagram in Appendix~\ref{app:calibration_bins}.} The score margin serves as a confidence signal: empirical accuracy rises with the predicted margin $m = |s(A){-}s(B)|$, from ${\sim}0.46$ at $m{\le}0.13$ to ${\sim}0.97$ at $m{\ge}2.64$ (Appendix~\ref{app:calibration_bins}).

\paragraph{Per-dataset contribution.}
Every training dataset contributes (leave-one-out retrains, Table~\ref{tab:internal_per_dataset}). Removing a dataset costs $0.029$ (MusicPrefs) to $0.093$ (SongEval) of accuracy on its own test split, with the SongEval drop inflated by its high-discriminability gap-filtered pairs. AIME dominates the full-set drop ($-0.041$ on all $2{,}035$ pairs): it makes up $77\%$ of the test pairs and carries $12{,}480$ of the $17{,}554$ training pairs. Off-diagonal movements are small relative to single-seed noise, the Music Arena column especially. The same trade-off between training on all four datasets and leaving one out recurs on external metrics (Appendix~\ref{app:cmi_external_detail}).

\begin{table}[!ht]
\centering
\footnotesize
\setlength{\tabcolsep}{3pt}
\caption{TuneJury held-out test accuracy by training mix (rows) and test split (columns). \emph{Full} is the released checkpoint trained on all four datasets; each $-$X row retrains without dataset X. {\setlength{\fboxsep}{0pt}\colorbox{gray!15}{~Shaded~}}: each leave-out model on its excluded dataset's split (OOD). Music Arena cells are noisy at $n{=}20$.}
\label{tab:internal_per_dataset}
\begin{tabular}{lrccccc}
\toprule
& & \multicolumn{5}{c}{Test split} \\
\cmidrule(lr){3-7}
Training mix & Train pairs & AIME ($1{,}560$) & MusicPrefs ($206$) & Music Arena ($20$) & SongEval ($249$) & All ($2{,}035$) \\
\midrule
Full & $17{,}554$ & $0.674$ & $0.718$ & $0.800$ & $0.908$ & $0.709$ \\
$-$AIME & $5{,}074$  & \cellcolor{gray!15}$0.625$ & $0.689$ & $0.650$ & $0.920$ & $0.668$ \\
$-$MusicPrefs & $15{,}542$ & $0.672$ & \cellcolor{gray!15}$0.689$ & $0.700$ & $0.912$ & $0.703$ \\
$-$Music Arena & $16{,}983$ & $0.673$ & $0.704$ & \cellcolor{gray!15}$0.750$ & $0.908$ & $0.706$ \\
$-$SongEval & $15{,}063$ & $0.686$ & $0.718$ & $0.750$ & \cellcolor{gray!15}$0.815$ & $0.706$ \\
\bottomrule
\end{tabular}
\end{table}

\paragraph{Sanity check on edge inputs.}
TuneJury scores silence and noise well below the $-0.18$ mean reward of the $n{=}20$ MTG-Jamendo reference sample, and synthetic tones below or near it (Appendix~\ref{app:sanity}), supporting its use as a coarse dataset-curation filter when the threshold is calibrated against the user's reference music distribution.

\paragraph{Input ablation.}
Seven variants differ only in their input feature stack (full table in Appendix~\ref{app:feature_modality}). Each row is a single-seed retrain at seed $42$. The released checkpoint uses the full three-block stack (CLAP audio $+$ MERT $+$ CLAP text). Text-only input is near random ($0.515$), confirming that the signal is primarily audio-derived. The six audio-containing variants land within a tight $0.013$-band ($0.695$ to $0.708$ Overall), and within-band ordering is sensitive to seed at this margin. Both the released three-block stack and single-block CLAP audio sit inside this band, leaving downstream users flexibility in input scope.

\subsection{External evaluation: CMI-RewardBench}
\label{sec:eval_external}

\begin{table}[t]
\centering
\caption{Scoring-model comparison on CMI-RewardBench test splits. Train input codes: T~$=$~text, L~$=$~lyrics, R~$=$~reference audio, A~$=$~audio. \textbf{Bold}/\underline{underline} mark best/$2$nd per column among OOD entries. $(\mathit{italic})$ marks in-distribution cells (CMI-RM on MusicEval and CMI-Pref; MuQ-Eval-A1 on MusicEval), excluded from OOD ranking.}
\label{tab:head_to_head}
\footnotesize
\setlength{\tabcolsep}{3pt}
\renewcommand{\arraystretch}{0.95}
\begin{tabular}{llcccc}
\toprule
& & \multicolumn{2}{c}{Musicality SRCC} & \multicolumn{2}{c}{Pairwise accuracy} \\
\cmidrule(lr){3-4} \cmidrule(lr){5-6}
Model & Train & PAM & MusicEval & CMI-Pref & Music Arena \\
\midrule
PAM score~\cite{deshmukh2024pam}                  & A, zero-shot, MS-CLAP  & $0.6098$ & $0.6733$ & $0.6640$ & $0.6709$ \\
Audiobox-Aesthetics~\cite{tjandra2025audiobox}        & A, $4$-axis MOS        & $0.5370$ & $0.6240$ & $0.7160$ & $0.6739$ \\
SongEval-RM~\cite{yao2025songeval}                & A, $5$-axis MOS, MuQ-large & $\mathbf{0.6977}$ & $0.6949$ & $0.7240$ & $\mathbf{0.7388}$ \\
MuQ-Eval-A1~\cite{zhu2026muqeval}                 & A, $2$-axis MOS, MuQ-large & $0.4995$ & $(\mathit{0.8089})$ & $0.6600$ & $0.6761$ \\
CMI-RM~\cite{ma2026cmirewardbench}                     & TLRA, $+$110K pseudo, MuQ-MuLan & $0.6606$ & $(\mathit{0.8266})$ & $(\mathit{0.7820})$ & $\underline{0.7343}$ \\
\midrule
TuneJury (T$+$A)                            & TA, 17.5K, CLAP$+$MERT & $0.6100$ & $0.6687$ & $0.7140$ & $0.7194$ \\
TuneJury (A only)$^{\dagger}$               & TA, 17.5K, CLAP$+$MERT & $\underline{0.6731}$ & $0.6618$ & $0.7240$ & $0.7007$ \\
\midrule
\multicolumn{6}{l}{\emph{Design-space ablations}$^{\dagger\dagger}$ \emph{(diagnostic variants; Section~\ref{sec:applications} Mode~1--3 results use the released variant above)}} \\
TuneJury, $-$SE (T$+$A)        & TA, 15K, CLAP$+$MERT   & $0.6331$ & $\underline{0.7154}$ & $0.7120$ & $0.7149$ \\
TuneJury, $-$MA (T$+$A)        & TA, 17K, CLAP$+$MERT   & $0.6381$ & $0.7100$ & $\underline{0.7380}$ & $0.6910$ \\
TuneJury, $-$MP (T$+$A)        & TA, 15.5K, CLAP$+$MERT & $0.6238$ & $0.6539$ & $0.7180$ & $0.7000$ \\
TuneJury, MuQ (T$+$A)          & TA, 17K, MuQ-MuLan   & $0.6146$ & $\mathbf{0.7848}$ & $\mathbf{0.7680}$ & $0.7004$ \\
\bottomrule
\end{tabular}

\smallskip
{\footnotesize\raggedright MA, MP, SE = Music Arena, MusicPrefs, SongEval. $\dagger$~empty prompt at inference. $\dagger\dagger$~training-mix or encoder ablation. ``$-$X'' excludes dataset X; the MuQ row swaps the encoder to MuQ-MuLan-large. All TuneJury variants are item-disjoint from CMI-RewardBench MA. MuQ-Eval-A1 (post-dates benchmark): our runs of its Hugging Face checkpoint.}
\end{table}
\vspace{-0.8em}

We evaluate TuneJury on the four CMI-RewardBench~\cite{ma2026cmirewardbench} test splits. PAM~\cite{deshmukh2024pam} ($500$ clips) and MusicEval~\cite{liu2025musiceval} ($413$ clips) report musicality SRCC, while CMI-Pref (the preference test split, $500$ pairs) and CMI-RewardBench's $1{,}340$-pair Music Arena split report pairwise accuracy. With the deployed text+audio protocol (the prompt is fed to the text branch), TuneJury reaches $0.610$, $0.669$, $0.714$, and $0.719$, respectively (Table~\ref{tab:head_to_head}). Item-level disjointness from our training pool is verified for PAM, MusicEval, and CMI-Pref (Appendix~\ref{app:cmi_external_detail}), and the Music Arena split is item-disjoint by construction (bench-clean removal, Section~\ref{sec:tunejury}).

\paragraph{Head-to-head with prior reward models.}
Table~\ref{tab:head_to_head} compares TuneJury against five prior baselines (in-distribution cells flagged in the caption).

\emph{Matched setup (no pseudo-label augmentation).} At ${\sim}17.5$\,K human-rated pairs and ${\sim}2.8$\,M trainable parameters, TuneJury exceeds CMI-RewardBench's own no-pseudo CMI-RM ablation~\cite{ma2026cmirewardbench} ($6{,}647$ human pairs, same TLRA inputs as full CMI-RM, ${\sim}30$\,M params) by ${+}0.17$ on PAM SRCC (musicality and alignment averaged per CMI-RewardBench's Table~$4$ reporting convention; TuneJury Mean SRCC $0.43$ vs.\ Scratch$+$Both $0.26$, where Scratch$+$Both is CMI-RewardBench's random-initialization ablation trained on CMI-Pref $+$ MusicEval without pseudo-label pretraining) and ${+}0.17$ on MusicEval musicality SRCC ($0.67$ vs.\ $0.50$). The $17.5$\,K vs $6.6$\,K data-volume difference is part of the design point: we train on the four open human-labeled sources without pseudo-augmentation. A matched-volume sub-sample comparison is left to future work. Against the two leaders (SongEval-RM and the pseudo-augmented full CMI-RM), the released text+audio TuneJury sits within $1$--$2$\,pp on CMI-Pref and CMI-RewardBench Music Arena. At matched single-input deployment (A-only), TuneJury leads PAM score on PAM by ${\sim}0.06$ SRCC and matches it within $0.02$ SRCC on MusicEval.

\emph{Design-space ablations.} The ablation rows of Table~\ref{tab:head_to_head} isolate three factors behind the residual gap to the leaders.
\textbf{(i)~Encoder:} swapping LAION-CLAP$+$MERT for MuQ-MuLan-large (a music-text contrastive encoder from the same MuQ family that SongEval-RM and CMI-RM rely on) lifts CMI-Pref to $0.7680$, the highest among OOD entries in the table.
\textbf{(ii)~Training-mix breadth:} leave-MA-out reaches $0.7380$ on CMI-Pref (above the released $0.7140$), a per-axis trade-off the broader four-source mix accepts to support Section~\ref{sec:applications} Modes~1--3.
\textbf{(iii)~Design point vs.\ optimum:} the MuQ-encoder variant leads on MusicEval and CMI-Pref, and the mix-controlled probe (Section~\ref{sec:tunejury}, Appendix~\ref{app:cmi_external_detail}) confirms this is a genuine encoder-axis gain, not a mix artifact: with the mix held fixed, MuQ matches or beats CLAP$+$MERT on four of five OOD axes. We release the MuQ-MuLan checkpoint alongside CLAP$+$MERT and read the gain as evidence that the head template is encoder-agnostic. CLAP$+$MERT (\texttt{tunejury.pt}) stays the default: it is trained on the full four-dataset mix behind every Mode~1--3 application, whereas the MuQ point used the reduced probe mix. On PAM, SongEval-RM (the MuQ-encoded $5$-axis MOS regressor) sits ${\sim}0.025$ SRCC above the best TuneJury variant. An AIME held-out comparison, in-distribution for TuneJury and therefore a sanity check rather than a head-to-head claim, appears in Appendix~\ref{app:aime_baselines}.

\paragraph{Text branch effect by prompt format.}
The text branch helps on the split whose prompts match our training distribution, hurts on the most mismatched split, and moves within noise elsewhere. Comparing the two TuneJury rows in Table~\ref{tab:head_to_head} (T$+$A vs.\ A only) isolates this effect (per-split $\Delta$ in Appendix~\ref{app:cmi_external_detail}): the text branch helps on CMI-RewardBench Music Arena (${+}1.87$\,pp), whose prompts share the live-arena style of our Music Arena training source, and hurts on PAM (${-}0.063$ SRCC), whose prompts are post-hoc captions of existing audio. Outside the live-battle style, the zero-vector empty-prompt protocol of Section~\ref{sec:tunejury} can be a safer default. On the $2{,}035$-pair internal held-out test the contribution is not statistically distinguishable from chance. T$+$A and A-only differ on $169$ pairs ($8.3\%$), with T$+$A correct on $89$ and A-only on $80$ (McNemar $\chi^2 = 0.48$, $p \approx 0.49$).

\paragraph{Musicality vs.\ text-alignment asymmetry.}
TuneJury correlates much more strongly with PAM's musicality MOS than with its text-alignment MOS (SRCC $0.610$ vs.\ $0.253$). This partly reflects the training labels: arena-style A~vs.~B preferences (Music Arena, AIME) collapse multiple rater considerations into one winner, MusicPrefs excludes alignment from its annotation~\cite{huang2025musicprefs}, and SongEval rates aesthetic axes only. A scalar trained on these labels learns a single composite axis, and PAM shows that composite leans toward musicality rather than alignment. Whether this asymmetry reflects raters weighting musicality more heavily, or musicality being easier to read from audio embeddings than alignment is from joint text-audio embeddings, remains unidentifiable from these collapsed labels.

\paragraph{Per-system PAM ordering: AI ordering recovered, real music underrated.}
PAM scores $100$ clips from each of $4$ TTM systems and a real-music reference. On the four TTM systems, TuneJury's per-system mean recovers PAM's musicality MOS order exactly: MusicGen-large $\succ$ MusicGen-melody (melody-conditioned variant)~\cite{copet2023musicgen} $\succ$ AudioLDM2-music~\cite{liu2024audioldm2} $\succ$ MusicLDM~\cite{chen2024musicldm}. The only mismatch is the real-music subsystem: $1$st by PAM musicality MOS but $3$rd by TuneJury mean (below both MusicGen variants, above AudioLDM2-music and MusicLDM). The all-system SRCC against PAM musicality MOS is $0.70$ on $n{=}5$ systems (AI-only SRCC ${+}1.00$ on the four TTM systems). With so few systems, we treat this as a descriptive ordering rather than a significance test. Two factors contribute to this real-music underrating. (i)~The real vs.\ AI calibration signal in the training mix is sparse: only AIME~\cite{grotschla2025aime} contains real-music pairs (via its MTG-Jamendo~\cite{bogdanov2019mtg} subset), and even there AIME crowdworkers preferred real over AI only weakly above chance (MTG-Jamendo real-audio baseline wins ${\sim}59\%$ of its comparisons in Appendix~\ref{app:per_system}). The remaining training pairs are AI vs.\ AI, so the learned scalar has limited supervision for scaling real music relative to AI. (ii)~The preference votes TuneJury learns from weigh factors beyond musicality (e.g., genre or instrumentation preferences) that PAM's musicality MOS does not, so the two scores partly measure different things.
\section{Applications: Selection, Inference-Time Optimization, and Post-Training}
\label{sec:applications}

Beyond benchmark accuracy, we exercise TuneJury as a preference-alignment signal in three downstream applications: inference-time selection, reward-driven latent optimization, and expert-iteration post-training. Each application tests whether the same frozen TuneJury can align a music generation pipeline with human preferences. Concretely, Mode~1 (best-of-$N$ selection) ranks frozen-backbone candidates by reward, Mode~2 (reward-driven latent optimization) backpropagates through DITTO~\cite{novack2024ditto} into the noise latents, and Mode~3 (expert iteration) fine-tunes the backbone on its own top-reward decile.

Mode~1 sweeps four frozen open-weights backbones spanning three architecture families. \href{https://huggingface.co/facebook/musicgen-medium}{MusicGen-medium} and \href{https://huggingface.co/facebook/musicgen-large}{MusicGen-large}~\cite{copet2023musicgen} are $1.5$\,B and $3.3$\,B autoregressive transformers. \href{https://huggingface.co/cvssp/audioldm2-music}{AudioLDM2-music}~\cite{liu2024audioldm2} is $1.1$\,B latent diffusion. \href{https://huggingface.co/ACE-Step/acestep-v15-turbo-continuous}{ACE-Step v$1.5$ Turbo Continuous}~\cite{aceStepV15} is a $2.4$\,B DiT with a continuous-audio latent decoder, released after our 2026-01 Music Arena cutoff (its outputs are unseen during training). Mode~2 and Mode~3 backbones are introduced in their respective subsections.

Audio samples from all three modes, with per-sample TuneJury scores, are available at the released listening demo (\href{https://huggingface.co/spaces/TuneJury/tune-jury-demo}{Hugging Face Space \texttt{TuneJury/tune-jury-demo}}).

\paragraph{Evaluation setup.}
Mode~1 and Mode~3 evaluate on \emph{SDD-$100$}: a $100$-prompt internal subset drawn from the $706$-entry Song Describer Dataset~\cite{manco2023sdd} (not an official split), prefixed with ``\texttt{high quality instrumental music, }''. We hold $n{=}100$ as the per-cell prompt budget: Mode~1 sweeps $N \in \{1,2,4,8,16,32\}$ on four backbones ($24$ generation cells, $12{,}800$ candidate generations in total), and Mode~3 evaluates baseline and post-trained checkpoints under the learning-rate sweep and multi-round probe (Appendix~\ref{app:mode3_ablations}). Mode~2 evaluates TangoFlux on the full $100$-prompt set and SAO-small on a $30$-prompt subset (Section~\ref{sec:apps_mode2}). We focus on instrumental music: the prompt prefix is applied to all four backbones, and ACE-Step Turbo Continuous (the only one with a separate lyric input) additionally receives an empty lyric.

\paragraph{Distributional metric choice.}
Modes~1--3 report on the same three axes: a distributional fidelity metric against SDD-$706$ (the full dataset's $706$ MTG-Jamendo audio tracks, used as the reference distribution)~\cite{manco2023sdd, bogdanov2019mtg}, the CLAP score (text-audio cosine similarity)~\cite{wu2023clap}, and mean TuneJury reward (Mode~1 in Figure~\ref{fig:mode1_bon_sweep}; Mode~2 and Mode~3 in Table~\ref{tab:apps_mode2_3}). For Mode~1 we report FAD-CLAP at $n{=}100$ per cell (Appendix~\ref{app:bon_full} adds FAD-MERT~\cite{li2024mert} and MAD~\cite{huang2025musicprefs}). For Mode~2 and Mode~3 we report MAD~\cite{huang2025musicprefs} on $1024$-d MERT embeddings, defined as $-\ln(\text{MAUVE})$~\cite{pillutla2021mauve} between $k$-means cluster occupancy histograms of the two sets, with range $[0,\infty)$ (lower means closer to the reference, aligning with FAD's direction). MAD compares cluster histograms rather than an empirical covariance, so it remains usable at the $n{=}30$ SAO-small cell, where a $512$-d covariance estimate from $30$ samples makes FAD-CLAP unreliable. Scaling $n$ to stabilize FAD-CLAP is prohibitive at Mode~2's per-prompt full-sampler backpropagation cost.

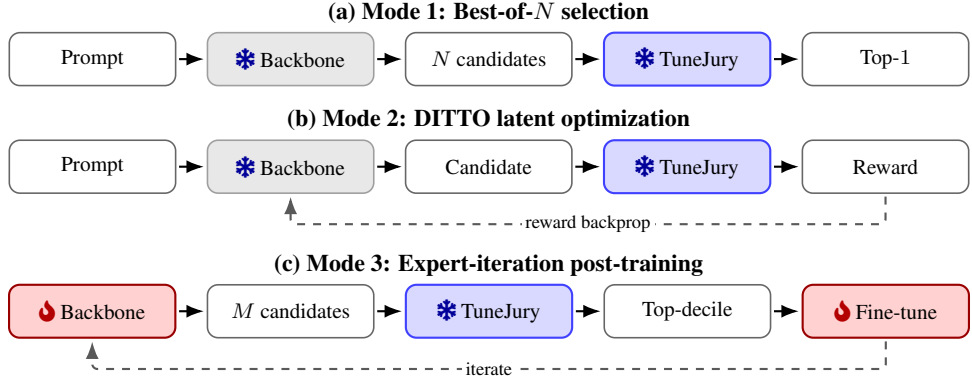
\begin{figure}[!t]
\centering
\tikzset{
    base/.style={draw, rounded corners, font=\footnotesize, align=center, inner sep=2pt, minimum height=7mm, minimum width=22mm, line width=0.6pt},
    neutral/.style={base, fill=white, draw=black!60},
    frozen/.style={base, fill=gray!18, draw=gray!75},
    trainable/.style={base, fill=red!18, draw=red!60!black, line width=0.8pt},
    reward/.style={base, fill=blue!15, draw=blue!75, line width=0.8pt},
    arr/.style={->, >=Latex, thick, shorten >=1pt, shorten <=1pt},
    feedback/.style={arr, dashed, draw=black!65}
}

\newcommand{\snow}{{\color{blue!60!black}\faSnowflake}}
\newcommand{\fire}{{\color{red!70!black}\faFire}}

\textbf{\small (a) Mode 1: Best-of-$N$ selection}\\[2pt]
\begin{tikzpicture}[node distance=4mm]
\node[neutral]                  (p)   {Prompt};
\node[frozen,  right=of p]      (bb)  {\snow~Backbone};
\node[neutral, right=of bb]     (c)   {$N$ candidates};
\node[reward,  right=of c]      (tj)  {\snow~TuneJury};
\node[neutral, right=of tj]     (out) {Top-$1$};
\draw[arr] (p) -- (bb);
\draw[arr] (bb) -- (c);
\draw[arr] (c) -- (tj);
\draw[arr] (tj) -- (out);
\end{tikzpicture}

\vspace{4pt}
\textbf{\small (b) Mode 2: DITTO latent optimization}\\[2pt]
\begin{tikzpicture}[node distance=4mm]
\node[neutral]                  (p)   {Prompt};
\node[frozen,  right=of p]      (bb)  {\snow~Backbone};
\node[neutral, right=of bb]     (c)   {Candidate};
\node[reward,  right=of c]      (tj)  {\snow~TuneJury};
\node[neutral, right=of tj]     (r)   {Reward};
\draw[arr] (p) -- (bb);
\draw[arr] (bb) -- (c);
\draw[arr] (c) -- (tj);
\draw[arr] (tj) -- (r);
\draw[feedback, rounded corners=2pt] (r.south) -- ++(0,-4mm) -| (bb.south);
\node[font=\scriptsize, fill=white, inner sep=1.5pt] at ($(bb.south)!0.5!(r.south) + (0,-4mm)$) {reward backprop};
\end{tikzpicture}

\vspace{2pt}
\textbf{\small (c) Mode 3: Expert-iteration post-training}\\[2pt]
\begin{tikzpicture}[node distance=4mm]
\node[trainable]                (bb)  {\fire~Backbone};
\node[neutral, right=of bb]     (g)   {$M$ candidates};
\node[reward,  right=of g]      (tj)  {\snow~TuneJury};
\node[neutral, right=of tj]     (f)   {Top-decile};
\node[trainable, right=of f]    (sft) {\fire~Fine-tune};
\draw[arr] (bb) -- (g);
\draw[arr] (g) -- (tj);
\draw[arr] (tj) -- (f);
\draw[arr] (f) -- (sft);
\draw[feedback, rounded corners=2pt] (sft.south) -- ++(0,-4mm) -| (bb.south);
\node[font=\scriptsize, fill=white, inner sep=1.5pt] at ($(bb.south)!0.5!(sft.south) + (0,-4mm)$) {iterate};
\end{tikzpicture}

\caption{Three downstream applications sharing a frozen TuneJury reward signal. Gray marks the frozen backbone, blue TuneJury (always frozen), and red the trainable backbone (Mode 3). $N$, $M$, and the top-decile filter are user-chosen hyperparameters. We use $N{\in}\{1,2,4,8,16,32\}$ (Section~\ref{sec:apps_bon}) and $M{=}900$ with top-$90$ filter (Section~\ref{sec:apps_mode3}).}
\label{fig:modes_pipeline}
\end{figure}

\subsection{Mode 1: Inference-time best-of-\texorpdfstring{$N$}{N} selection}
\label{sec:apps_bon}

\begin{figure}[!ht]
\centering
\includegraphics[width=\linewidth]{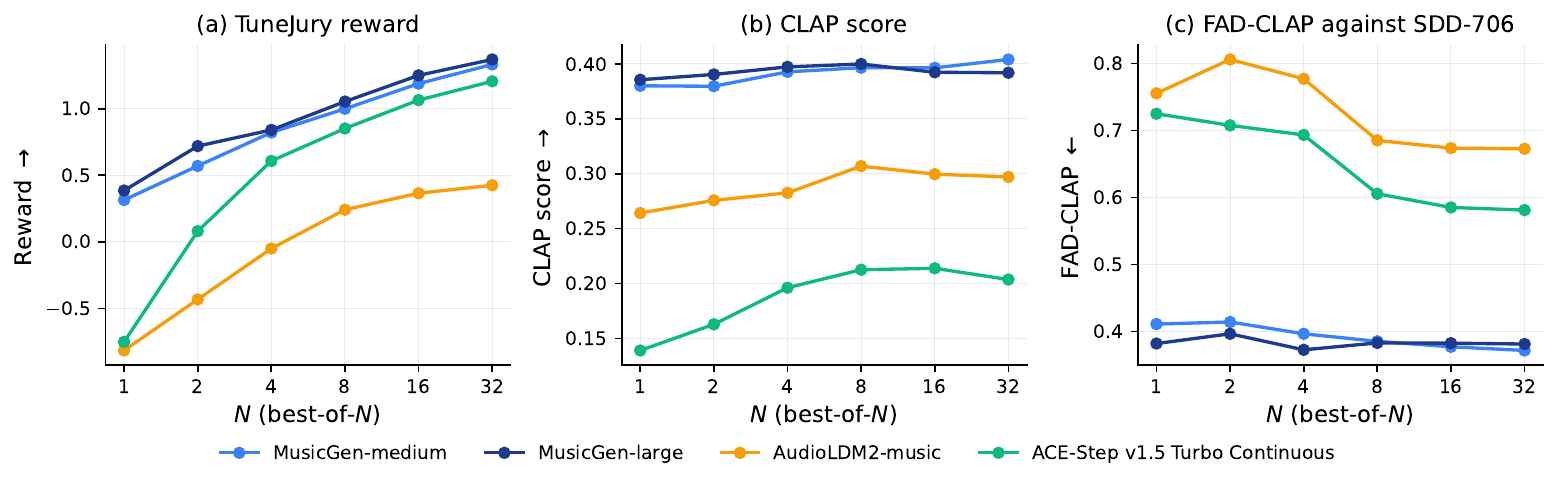}
\caption{Mode~1 best-of-$N$ sweep ($N \in \{1, 2, 4, 8, 16, 32\}$) on four frozen open-weights backbones
with the released bench-clean TuneJury as the selector.
\emph{(a)}~TuneJury reward. \emph{(b)}~CLAP score.
\emph{(c)}~FAD-CLAP against SDD-$706$. Reward is monotone in $N$ on
every backbone. CLAP score and FAD-CLAP improvements vary by backbone.
Per-backbone exact values for all five metrics (FAD-CLAP, CLAP score, FAD-MERT, MAD, Reward) at every $N$ are in Table~\ref{tab:apps_mode1_bon_full}.}
\label{fig:mode1_bon_sweep}
\end{figure}

We generate $N \in \{1, 2, 4, 8, 16, 32\}$ candidates per prompt at each backbone's defaults (only the noise seed differs between candidates), score each candidate with TuneJury against the same prompt, and keep the Top-$1$. Results across all four backbones are reported in Figure~\ref{fig:mode1_bon_sweep}. Reward is strictly monotone in $N$ on every backbone. The per-doubling gain narrows from $[{+}0.178, {+}0.291]$ at $N{=}4{\to}8$ to $[{+}0.060, {+}0.144]$ at $N{=}16{\to}32$. All four backbones show decelerating per-doubling gain in this final step. AudioLDM2-music (the backbone with the lowest $N{=}1$ reward) saturates earliest with the smallest gain ($+0.060$). Per-doubling values per backbone are in Table~\ref{tab:apps_mode1_bon_full}.

\paragraph{Reward signal: audio-driven, with text-audio alignment as a byproduct.}
The released scalar is audio-driven rather than driven by text alignment (Section~\ref{sec:eval_internal}), yet Mode~1 improves alignment as a byproduct. The CLAP score rises with $N$ on every backbone (Appendix~\ref{app:mode1_extended}), so the TuneJury preference score and the CLAP score are positively correlated in the candidate distributions produced by Mode~1. The correlation is mediated by the training distribution and is not guaranteed to transfer, so for OOD prompts we recommend reporting a dedicated alignment metric alongside TuneJury. Mode~1's positive per-doubling gain through $N{=}32$ differs from CMI-RewardBench~\cite{ma2026cmirewardbench}'s reported best-of-$N$ saturation. Appendix~\ref{app:bon_monotone} attributes the difference to setup choices.

\paragraph{Distributional metrics disagree across encoders.}
Three distributional metrics against SDD-$706$ (FAD-CLAP, FAD-MERT, and MAD) produce three different per-backbone patterns as $N$ grows in Mode~1. FAD-CLAP improves at $N{=}4$ on three backbones (MusicGen-medium/large, ACE-Step Turbo Continuous) and worsens at $N{=}4$ on AudioLDM2-music before recovering to the sweep best at $N{=}32$ (full trajectory in Table~\ref{tab:apps_mode1_bon_full}). FAD-MERT moves the opposite way at $N{=}4$ on three of those four. It worsens on the two MusicGen variants where FAD-CLAP improves, and improves on AudioLDM2-music where FAD-CLAP worsens. MAD on MERT, despite sharing an encoder with FAD-MERT, ends below its $N{=}1$ value on all four backbones (lower MAD means closer to SDD-$706$; Table~\ref{tab:apps_mode1_bon_full}). Two of the four reach their minimum before $N{=}32$ (AudioLDM2-music at $N{=}8$, ACE-Step Turbo Continuous at $N{=}16$) and rebound by $N{=}32$. The two MusicGen variants reach their minimum at $N{=}32$. A diversity probe (Appendix~\ref{app:mode1_diversity}) rules out mode collapse on the two rebounding backbones. Their top-$1$ picks spread more at higher $N$, not less. The rebound therefore reflects distributional drift, a partial inference-time analog of the Mode~3 reward-fidelity trade-off (Section~\ref{sec:apps_mode3}) that surfaces on the two backbones with the largest reward headroom. Encoder choice (LAION-CLAP vs.\ MERT) and divergence measure (FAD vs.\ MAD) each change the per-backbone reading. Practitioners evaluating TTM systems with a text-aligned reward should triangulate FAD-CLAP, FAD-MERT, and MAD rather than rely on a single distributional metric. On Mode~3 expert iteration, the drift appears on MAD (a step pattern with $10^{-6}$ and $5{\times}10^{-6}$ essentially tied and $10^{-5}$ rising further) while the CLAP score stays approximately flat (Section~\ref{sec:apps_mode3}), so the two side metrics disagree there as well.

\paragraph{Lift tracks the $N{=}1$ reward headroom.}
Top-$1$ best-of-$N$ reward cannot decrease in $N$ by construction, so the empirical claim is not mere monotonicity but the \emph{shape} of the marginal gain (Appendix~\ref{app:mode1_extended}). The lift tracks where the $N{=}1$ distribution sits relative to the high-reward tail: backbones with the largest reward gap at $N{=}1$ (ACE-Step Turbo Continuous, AudioLDM2-music) gain the most by $N{=}4$, while the MusicGen variants, already close to the tail, gain less.

\subsection{Mode 2: Inference-time latent optimization}
\label{sec:apps_mode2}

\paragraph{DITTO protocol.}
We apply DITTO-style optimization~\cite{novack2024ditto} to two backbones, \href{https://huggingface.co/stabilityai/stable-audio-open-small}{SAO-small}~\cite{novack2025saoarc} and \href{https://huggingface.co/declare-lab/TangoFlux}{TangoFlux}~\cite{hung2025tangoflux}, with TuneJury as the reward and the Mode~1 prompt prefix. We run both samplers at $8$ denoising steps, freeze the base weights, and optimize only the initial noise latents. Each of $5$ iterations runs the full chain, scores the output with TuneJury, and backpropagates through all $8$ steps to update the noise toward higher reward (Adam optimizer~\cite{kingma2015adam}, learning rate $0.05$). %We omit AudioLDM2-music: its $50$-step denoising trajectory exceeds the full-backprop budget on our hardware.

\paragraph{Side metrics split by baseline headroom.}
DITTO lifts mean TuneJury reward on both backbones (Table~\ref{tab:apps_mode2_3}, top), and the lift is larger on the backbone whose baseline reward is lower (TangoFlux from $-0.978$ vs.\ SAO-small from $+0.159$). SAO-small is evaluated on a $30$-prompt SDD-$100$ subset and TangoFlux on the full $100$-prompt set (a reproducibility constraint of the SAO-small release at the time of writing; Appendix~\ref{app:sao_caveat}). The two side metrics split per backbone. On TangoFlux, MAD against SDD-$706$ drops sharply (${-}2.214$) and the CLAP score rises (${+}0.043$): DITTO pulls a low-reward backbone toward audio that is closer to SDD-$706$ and better aligned with the text prompt, a win-win pattern with no visible reward exploitation. On SAO-small, both side metrics regress (MAD ${+}0.500$, CLAP score ${-}0.007$): the baseline already sits at near-zero reward (${+}0.159$), and DITTO's reward gain (${+}0.245$) comes at the cost of distributional and alignment drift, the classic three-axis reward-exploitation pattern~\cite{gao2023scaling}. The learning-rate sweep in Section~\ref{sec:apps_mode3} demonstrates the same reward-fidelity tension under more controlled conditions.

\begin{table}[!ht]
\centering
\caption{Mode~2 (DITTO, top) and Mode~3 (expert iteration on FluxAudio-S, bottom). Each block lists the baseline and its post-optimization rows. \emph{Reward} is mean TuneJury reward, \emph{MAD}~\cite{huang2025musicprefs} is $-\ln(\text{MAUVE})$ on $1024$-d MERT embeddings against SDD-$706$ (lower means closer; protocol in Appendix~\ref{app:reproducibility}), and \emph{Win} counts prompts with a reward increase. Parentheses give the change from the baseline, computed before rounding. SAO-small's MAD rise with slightly lower CLAP score and Mode~3's MAD rise are the reward-fidelity trade-off of Section~\ref{sec:apps_mode3}, not failure modes.}
\label{tab:apps_mode2_3}
\footnotesize

\emph{Mode 2 (DITTO; SAO-small at $n{=}30$, TangoFlux at $n{=}100$)}\\[1pt]
\begin{tabular*}{\linewidth}{@{\extracolsep{\fill}}lcccc@{}}
\toprule
Model & Reward$\uparrow$ & MAD$\downarrow$ & CLAP score$\uparrow$ & Win \\
\midrule
SAO-small ($340$\,M) & $+0.159$ & $1.070$ & $0.1961$ & -- \\
\quad $+$ DITTO & $+0.404$ (${+}0.245$) & $1.570$ (${+}0.500$) & $0.1886$ (${-}0.007$) & $19/30$ \\
TangoFlux ($515$\,M) & $-0.978$ & $4.263$ & $0.1501$ & -- \\
\quad $+$ DITTO & $+0.578$ (${+}1.557$) & $2.048$ (${-}2.214$) & $0.1933$ (${+}0.043$) & $100/100$ \\
\bottomrule
\end{tabular*}

\vspace{6pt}

\emph{Mode 3 (expert iteration, SDD-$100$, FluxAudio-S backbone; learning-rate sweep, single round)}\\[1pt]
\begin{tabular*}{\linewidth}{@{\extracolsep{\fill}}lcccc@{}}
\toprule
Checkpoint & Reward$\uparrow$ & MAD$\downarrow$ & CLAP score$\uparrow$ & Win \\
\midrule
FluxAudio-S ($120$\,M) & $-0.262$ & $1.758$ & $0.0921$ & -- \\
\quad lr $10^{-6}$ (conservative) & $-0.096$ (${+}0.166$) & $2.051$ (${+}0.293$) & $0.1109$ (${+}0.019$) & $67/100$ \\
\quad lr $5{\times}10^{-6}$ & $+0.107$ (${+}0.369$) & $2.041$ (${+}0.284$) & $0.1195$ (${+}0.027$) & $73/100$ \\
\quad lr $10^{-5}$ (aggressive) & $+0.154$ (${+}0.416$) & $2.427$ (${+}0.669$) & $0.1155$ (${+}0.023$) & $75/100$ \\
\bottomrule
\end{tabular*}
\end{table}

\subsection{Mode 3: Expert-iteration post-training as a Pareto-frontier stress test}
\label{sec:apps_mode3}

Mode~3 post-trains the backbone weights themselves against TuneJury reward, using the publicly released FluxAudio-S checkpoint (${\sim}120$\,M rectified-flow DiT at $16$\,kHz; \href{https://huggingface.co/AndreasXi/MeanAudio/blob/main/fluxaudio_s_full.pth}{\texttt{fluxaudio\_s\_full.pth}} from the MeanAudio release~\cite{li2025meanaudio}) under expert iteration~\cite{anthony2017expertiteration, singh2024rest} on the model's own outputs. Each round generates $900$ candidates ($9$ noise seeds per SDD-$100$ prompt), scores them with TuneJury, retains the top reward decile ($90$ samples), and fine-tunes on those $90$ alone for $5$\,K iterations. No external data is mixed in at the fine-tune step, so the only training signal is the self-filtered expert set on top of the FluxAudio-S pretraining prior. We choose expert iteration over diffusion-side policy gradient (DDPO~\cite{black2024ddpo}, GRPO~\cite{shao2024deepseekmath}) because it is offline and model-agnostic: the loop requires only sampling, scoring, filtering, and supervised fine-tuning, with no online reinforcement learning through the denoising chain. Policy gradient instead must modify the sampler to track per-step action log-probabilities. Full hyperparameters and the inference configuration appear in Appendix~\ref{app:mode3_ablations}.

\paragraph{Learning-rate sweep traces a reward-fidelity trade-off.}
We frame Mode~3 as a Pareto-frontier stress test, mapping the reward-fidelity trade-off across three fine-tune learning rates ($10^{-6} / 5{\times}10^{-6} / 10^{-5}$, single round each, all other hyperparameters fixed; Table~\ref{tab:apps_mode2_3}, bottom). Reward lift grows monotonically with the learning rate (${+}0.166 \to {+}0.369 \to {+}0.416$). MAD against SDD-$706$ shows a step pattern: $10^{-6}$ and $5{\times}10^{-6}$ are essentially tied (${+}0.293$ and ${+}0.284$) and $10^{-5}$ rises noticeably further (${+}0.669$). The CLAP score stays approximately flat at a small positive offset (${+}0.019$ / ${+}0.027$ / ${+}0.023$), so the drift is distributional rather than a loss of text alignment. The pairing of reward gains with MAD drift is the classic reward-exploitation signature (a form of Goodhart's law~\cite{gao2023scaling}).

\paragraph{Drift is structural under instance-level reward optimization.}
TuneJury is an instance-level scalar with no reference-distribution term, so maximizing it under fine-tuning leaves no penalty for the backbone drifting off the SDD-$706$ manifold. Among the swept rates, $5{\times}10^{-6}$ is the most favorable trade-off. It more than doubles the reward lift over $10^{-6}$ at essentially the same MAD cost, and going further to $10^{-5}$ adds the remaining $0.047$ reward gain at more than twice the MAD penalty (we do not claim this is the global Pareto optimum, only the best of the three swept points). In a multi-round expert-iteration probe at learning rate $10^{-6}$, the reward collapses round over round and MAD drifts further (dropping below the baseline by the third round; Appendix~\ref{app:mode3_ablations}), consistent with the trade-off stemming from the objective itself rather than from the $10^{-6}$ single-round point alone.

\paragraph{Reward exploitation is independent evidence of a real preference signal.}
The reward-exploitation pattern surfacing under TuneJury optimization is independent evidence that TuneJury behaves like a real preference-alignment signal. Neither random noise nor a metric trivially aligned with MAD or the CLAP score would produce this consistent divergence between reward and distributional fidelity, which is the empirical signature of reward hacking on a meaningful but imperfect proxy~\cite{skalse2022defining}. Three concrete patches against the trade-off: pick the most favorable swept rate ($5{\times}10^{-6}$ here), anchor the fine-tune set with held-out external audio, or fold a distributional or alignment side metric into the expert filter.

\section{Discussion}
\label{sec:discussion}

\paragraph{Encoder choice carries more OOD lift than training-mix breadth.}
Holding the MLP head template and training mix fixed, swapping LAION-CLAP$+$MERT for MuQ-MuLan-large matches or beats the leave-MA-out CLAP$+$MERT baseline on four of five OOD axes at half the input dimensionality (single-seed probe at seed $42$; Appendix~\ref{app:cmi_external_detail}, ``Encoder swap probe''). At the ${\sim}17.5$\,K human-rated pair scale, the encoder swap yields larger OOD lift than the leave-one-out training-mix sweep we ran in the same appendix.

\paragraph{SongEval's gap filter inflates internal accuracy and degrades external musicality SRCC.}
SongEval's ${\ge}0.5$ mean-gap filter selects for high-discriminability pairs, inflating internal accuracy (Table~\ref{tab:internal_per_dataset}) and degrading external PAM and MusicEval SRCC (Table~\ref{tab:training_mix}). We retain SongEval in the released mix for per-dataset coverage. A more principled fix would be the mixed-supervision design in Open directions~(i).

\paragraph{TuneJury as a capability proxy for text-to-music systems.}
The score distribution a TTM system produces against TuneJury can serve as a quick capability proxy. On held-out test splits, per-system reward ranking matches per-system human win rate at $\rho{=}{+}0.98$ on AIME and $\rho{=}{+}0.96$ on MusicPrefs (in-distribution at the dataset level; Appendix~\ref{app:per_system}). Combined with the Mode~3 reward lift (baseline ${-}0.262$ to ${+}0.154$ at the aggressive learning rate), developers get an inexpensive early diagnostic on backbone choice and post-training headroom.

\paragraph{Limitations.}
TuneJury's calibration and rank correlations depend on the four-dataset mix and LAION-CLAP$+$MERT stack. (i)~Real vs.\ AI calibration signal is sparse (only AIME's MTG-Jamendo subset), so the per-system PAM diagnostic shows real music underrated relative to the AI ordering. (ii)~Vocal-music coverage is weak (mainly Music Arena and SongEval). (iii)~Arena clips ($10$--$30$\,s typical) and SongEval full tracks (median ${\sim}3.4$\,min) differ in length. Inference time-averages, so long-form within-song variation is lost. (iv)~Calibration bin boundaries (Appendix~\ref{app:calibration_bins}) are mix-specific. (v)~TuneJury is trained on pre-2026-02 Music Arena. Agreement drops on post-cutoff splits to ${\sim}0.54$ (Feb--Mar 2026) and ${\sim}0.64$ raw (April~2026). The drop partly reflects a pre-cutoff label-noise ceiling rather than pure model failure (intrinsic-difficulty decomposition, Appendix~\ref{app:postcut_decomposition}). Anchor calibration (below) recovers the new-system OOD component.

\paragraph{Anchor calibration.}
We fit a Bradley--Terry~\cite{bradley1952rank} per-system bias term on top of the frozen TuneJury score, holding one in-distribution system at $\beta{=}0$ for identifiability. With ${\sim}100$ post-cutoff calibration pairs, the procedure recovers ${\sim}5$\,pp of agreement without retraining, and the anchor at $K{=}10$ already matches a from-scratch retrain at $K{=}250$ (Figure~\ref{fig:ood_scaling}, Appendix~\ref{app:anchor_calibration}; code in \href{https://github.com/yonghyunk1m/TuneJury/tree/main/applications/anchor_calibration}{\texttt{applications/anchor\_calibration}}). The recovery is slice-dependent: Feb--Mar gains substantially, while April is already near the label-noise ceiling. Anchor calibration is therefore a targeted patch against the specific generator slice that drives the OOD drop, not a generic monthly refresh.

\paragraph{Open directions.}
Three directions follow from the released artifact.
\begin{enumerate*}[label=(\roman*),leftmargin=*]
\item \emph{Mixed instance-level $+$ pairwise supervision with alignment-targeted training:}
treat SongEval as $5$-axis instance-level regression while retaining the pairwise objective for arena-style sources, and add an alignment-supervised head trained on per-axis MOS at scale. Our decomposition probe (Appendix~\ref{app:decomp_probe}) suggests an alignment-specific signal in the features. The probe head's partial SRCC (controlling for musicality) is still ascending at the upper limit of the ${\sim}900$-clip alignment-labeled MOS pool, separate from TuneJury's ${\sim}17.5$\,K-pair preference training. Pseudo-label augmentation is one way to extend that probe beyond the current pool and test whether the trend continues.
\item \emph{Scaling reward-driven post-training:}
replace Mode~3 expert iteration with GRPO~\cite{shao2024deepseekmath} on additional open-weights backbones (MusicGen, ACE-Step Turbo Continuous), and extend Mode~2 DITTO autograd to backbones beyond the two reported in Section~\ref{sec:apps_mode2}.
\item \emph{Vocal-music scope extension:}
a vocal-capable backbone for Mode~1, a vocal-music reference set for distributional metrics, and the real vs.\ AI calibration pairs needed to address limitation~(i).
\end{enumerate*}

\section{Conclusion}
\label{sec:conclusion}

We release \textbf{TuneJury}, an open, instance-level pairwise music reward model trained on human-rated pairs from four open sources without pseudo-label augmentation. A small MLP head over frozen music-pretrained encoders generalizes to held-out test pairs and out-of-distribution benchmarks, staying competitive with the pseudo-augmented CMI-RM baseline on the latter. The same frozen scalar drives three downstream applications on open-weights backbones without per-mode tuning: inference-time best-of-$N$ selection, DITTO-style latent optimization, and expert-iteration post-training. The reward-fidelity trade-off under expert iteration is the classic reward-exploitation pattern, independent evidence that TuneJury behaves as a genuine preference-alignment signal rather than noise or a trivial restatement of side metrics. We additionally release \emph{anchor calibration}, a post-hoc, per-system Bradley--Terry calibration that adapts TuneJury to new TTM systems at substantially better data efficiency than retraining (Appendix~\ref{app:anchor_calibration}). All artifacts (checkpoints, application pipelines, calibration code, listening demos, and pre-computed scores on seven open collections) are released and documented in Appendix~\ref{app:release_detail}.

\section*{Acknowledgements}
This work was supported by funding from Sony AI.

Yinghao Ma is a research student at the UKRI Centre for Doctoral Training in Artificial
Intelligence and Music, supported by UK Research and Innovation [grant number EP/S022694/1].
Yinghao Ma also acknowledges the support of Google PhD Fellowship. 

{
\small
\bibliography{references}
\bibliographystyle{unsrtnat}
}

\appendix
\section*{Notation Used Throughout the Appendix}

The four training datasets are abbreviated as MA (Music Arena~\cite{kim2025musicarena}), MP (MusicPrefs~\cite{huang2025musicprefs}), AIME~\cite{grotschla2025aime} (no shorter form), and SE (SongEval~\cite{yao2025songeval}). Other recurring abbreviations: SRCC (Spearman rank correlation coefficient), ECE (expected calibration error), OOD (out-of-distribution), FAD-CLAP and FAD-MERT (FAD~\cite{kilgour2019fad} computed using LAION-CLAP-Music and MERT-v$1$-$330$M embeddings, respectively, against the SDD-$706$ reference), MAD (MAUVE Audio Divergence~\cite{huang2025musicprefs}, defined in Section~\ref{sec:applications}), CLAP score (text-audio cosine similarity~\cite{wu2023clap}), CMI-Pref (the CMI-RewardBench preference test split), and Mode~1 / Mode~2 / Mode~3 (the three downstream applications of Section~\ref{sec:applications}).

\section{Calibration: Reliability Diagram and Bins}
\label{app:calibration_bins}

\begin{figure}[h]
\centering
\includegraphics[width=\linewidth]{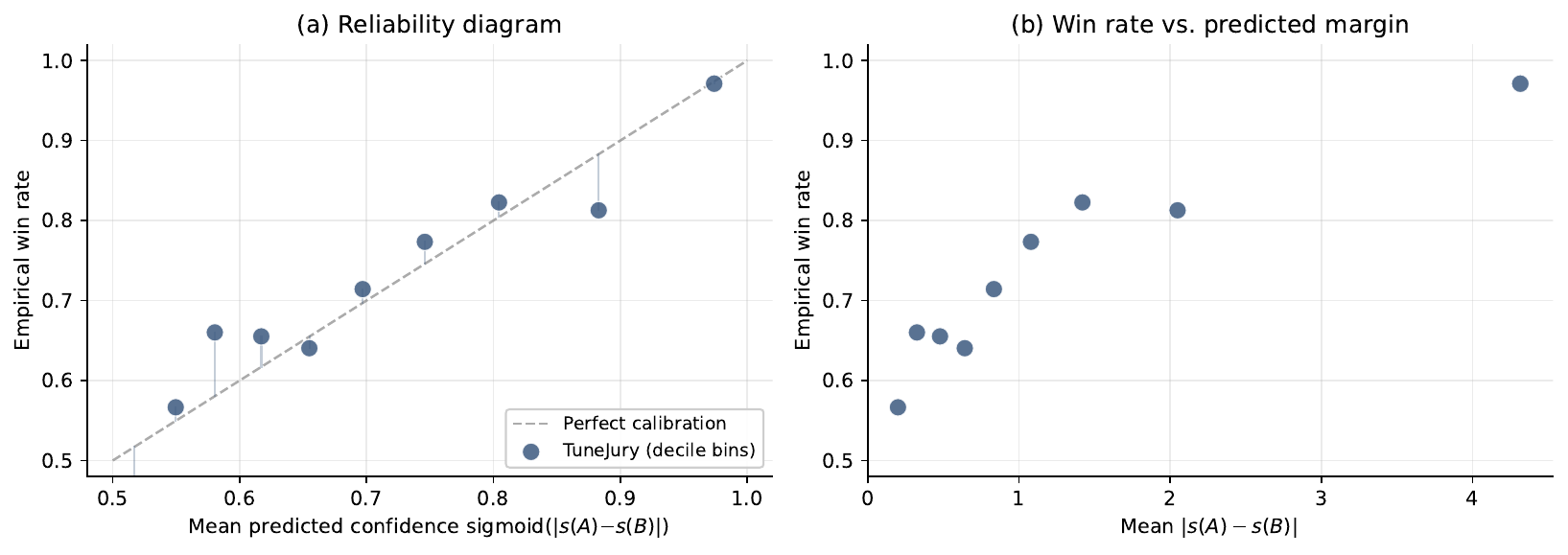}
\caption{TuneJury calibration on $n{=}2{,}035$ held-out test pairs (ties excluded). \emph{(a)} Reliability diagram: predicted confidence tracks win rate along $y{=}x$ (pairwise accuracy $0.7086$, ECE $0.0339$). \emph{(b)} Win rate vs.\ predicted margin $m{=}|s(A){-}s(B)|$, rising from ${\sim}0.46$ at $m{\le}0.13$ to ${\sim}0.97$ at $m{\ge}2.64$.}
\label{fig:calibration}
\end{figure}

Figure~\ref{fig:calibration} (reliability diagram) and Table~\ref{tab:calibration_bins} ($10$-bin decomposition) back the calibration claim and the margin-threshold decision rule of Section~\ref{sec:eval_internal}. Both are computed on the released TuneJury over the test partition of every training dataset (the Music Arena fold excludes any \texttt{battle\_uuid} in CMI-RewardBench's MA test split).

\begin{table}[h]
\centering
\caption{TuneJury reliability on $n{=}2{,}035$ held-out test pairs, binned by predicted margin $m{=}|s(A){-}s(B)|$. Win rate is non-decreasing across bins apart from small dips ($\le 0.02$ relative to the previous bin at bins $4$, $5$, and $9$). Bin~$1$ at $0.463$ reflects near-chance behavior below the $0.13$ margin threshold. Bin edges are deciles of the test-pair margin distribution. Log-loss $0.5547$ (overall metrics as in Figure~\ref{fig:calibration}).}
\label{tab:calibration_bins}
\footnotesize
\begin{tabular}{cccccc}
\toprule
Bin & $m$ range & $n$ & mean $m$ & Empirical win rate & Mean confidence \\
\midrule
$1$  & $[0.00, 0.13)$ & $203$ & $0.07$ & $0.463$ & $0.517$ \\
$2$  & $[0.13, 0.26)$ & $203$ & $0.20$ & $0.567$ & $0.550$ \\
$3$  & $[0.26, 0.40)$ & $203$ & $0.33$ & $0.660$ & $0.580$ \\
$4$  & $[0.40, 0.56)$ & $203$ & $0.48$ & $0.655$ & $0.617$ \\
$5$  & $[0.56, 0.73)$ & $203$ & $0.64$ & $0.640$ & $0.655$ \\
$6$  & $[0.73, 0.95)$ & $203$ & $0.83$ & $0.714$ & $0.697$ \\
$7$  & $[0.95, 1.22)$ & $203$ & $1.08$ & $0.773$ & $0.746$ \\
$8$  & $[1.22, 1.67)$ & $203$ & $1.42$ & $0.823$ & $0.804$ \\
$9$  & $[1.67, 2.64)$ & $203$ & $2.05$ & $0.813$ & $0.883$ \\
$10$ & $[2.64, 10.00]$ & $208$ & $4.32$ & $0.971$ & $0.974$ \\
\bottomrule
\end{tabular}
\end{table}

Bin~$10$'s wide range up to $10.00$ is dominated by SongEval test pairs. SongEval pairs are synthesized from random song pairings whose mean rating gap on the $5$ axes is $\ge 0.5$ (Section~\ref{sec:tunejury}), admitting larger quality gaps than arena-style pairs: $\max |s(A){-}s(B)|$ is $2.44$ on Music Arena, $3.93$ on MusicPrefs, $4.67$ on AIME, and $10.00$ on SongEval. The pairwise logistic loss pushes distinguishable pairs apart without bound, so extreme-margin pairs are an expected consequence of training on a mix with synthetic, high-contrast SongEval pairs.

\section{Adversarial Sanity Checks}
\label{app:sanity}

We probe TuneJury with synthetic and perturbed-music inputs. All probes are mono $10$\,s waveforms at $16$\,kHz scored under the released checkpoint's zero-vector empty-prompt protocol (Section~\ref{sec:tunejury}), with $20$ MTG-Jamendo clips as reference music. A non-empty prompt preserves the relative ordering (absolute scores shift but conclusions transfer), since the score is primarily audio-derived (input ablation: text-only $0.515$ near-random, Appendix~\ref{app:feature_modality}). Broadband noise sits furthest below music, with low-frequency sines also in the below-music regime. Spectrally colored noise (white $\to$ pink $\to$ brown) moves toward music as the spectrum becomes more low-frequency dominated, and harmonic structure on a low-frequency sine pulls the score upward.

\paragraph{Boundary inputs.}
Table~\ref{tab:sanity_boundary} lists scores for silence, three noise types (white at four amplitudes, pink, brown), isolated sine tones, and harmonic stacks. White noise sits in a flat band ($-3.9$ to $-4.6$) across amplitudes, confirming the score does not merely penalize low-energy inputs. Silence ($-1.05$) sits above broadband noise, likely because exact zeros drive the audio front-ends to constant activations while noise produces bin-varying activations.

\begin{table}[h]
\centering
\caption{TuneJury scores on adversarial / OOD inputs (empty-prompt protocol, $10$\,s waveforms at $16$\,kHz). The reference music row (last) defines the ``music regime'' (mean $-0.18\pm0.66$, range $[{-}1.39, {+}1.05]$) for visual contrast against synthetic inputs above. The $n{=}20$ MTG-Jamendo reference clips suffice because the per-clip variance is small relative to the score gap separating music from the worst synthetic inputs (${>}\,3$ score units to white noise).}
\label{tab:sanity_boundary}
\footnotesize
\begin{tabular}{lc}
\toprule
Input & Score \\
\midrule
Silence (zeros) & $-1.05$ \\
White noise, RMS $-60$ / $-40$ / $-20$ / $0$\,dBFS & $-3.90$ / $-4.03$ / $-3.97$ / $-4.59$ \\
Pink noise ($1/f$) & $-2.32$ \\
Brown noise ($1/f^2$) & $-1.99$ \\
Pure sine $110$ / $220$ / $440$ / $880$ / $1760$\,Hz & $-2.83$ / $-3.03$ / $-1.19$ / $-0.50$ / $-0.07$ \\
Harmonic stack $300$/$600$/$900$\,Hz & $-1.61$ \\
$A_2$ harmonic series ($110$--$660$\,Hz) & $-1.86$ \\
Reference music (MTG-Jamendo, $n{=}20$, $10$\,s) & $-0.18 \pm 0.66$, range $[-1.39, +1.05]$ \\
\bottomrule
\end{tabular}
\end{table}

\paragraph{Graded music-quality perturbations.}
Two perturbation ladders on the same $n{=}8$ MTG-Jamendo clip set: (i)~mix with white noise at SNR $\in \{40, 20, 10, 5, 0\}$\,dB and renormalize, (ii)~hard-clip to ratios $\{0.5, 0.1, 0.05, 0.02\}$ of peak amplitude and renormalize. Mean reward is strictly monotone along both axes (Table~\ref{tab:sanity_perturb}), and at the most aggressive clip ratios the score falls into the noise / synthetic regime of Table~\ref{tab:sanity_boundary}.

\begin{table}[h]
\centering
\caption{Mean TuneJury reward under graded perturbations on a fixed $n{=}8$ MTG-Jamendo clip set (empty-prompt protocol). Both ladders are strictly monotone. The clean baseline $-0.10$ is this $8$-clip set, distinct from the $n{=}20$ MTG-Jamendo reference in Table~\ref{tab:sanity_boundary}.}
\label{tab:sanity_perturb}
\footnotesize
\begin{tabular}{lcccccc}
\toprule
\textit{SNR (dB)}   & clean   & $40$    & $20$    & $10$    & $5$     & $0$     \\
Mean reward         & $-0.10$ & $-0.36$ & $-0.72$ & $-1.25$ & $-1.79$ & $-2.62$ \\
\midrule
\textit{Clip ratio} & clean   & $0.5$   & $0.1$   & $0.05$  & $0.02$  & n/a     \\
Mean reward         & $-0.10$ & $-0.31$ & $-1.48$ & $-1.90$ & $-2.43$ & n/a     \\
\bottomrule
\end{tabular}
\end{table}

\paragraph{Length sensitivity.}
Truncating the same $8$ clips to $\{1, 3, 5, 10\}$\,s yields mean reward $-0.75 \to -0.65 \to -0.46 \to -0.10$ (monotone increasing in available context up to $10$\,s, so cross-clip score comparisons are only meaningful at fixed length). Beyond $10$\,s, on a separate set of $8$ MTG-Jamendo tracks (each $\ge 60$\,s), mean reward peaks around $45$\,s before falling back at full track, reflecting diminishing benefit from the encoder's time-pooling beyond the training context. For fine-grained track-level ranking, a fixed-duration sliding-window rescore is the appropriate approach.

\paragraph{Per-segment discrimination probe.}
On a $50$\,s composite of five $10$\,s Jamendo clips with the central $10$\,s slot replaced by silence or $-20$\,dBFS white noise, a $10$\,s window at $5$\,s hop drops to the standalone score over the bad slot (silence $-1.05$, noise $-4.08$, both within Table~\ref{tab:sanity_boundary}'s adversarial-input regime) and stays at clean-music levels elsewhere. Segment-level discrimination at inference time is preserved despite TuneJury training on clip-level labels, supporting a sliding-window rescore.

\paragraph{Temporal-structure sensitivity.}
Time-reversing the same $8$ clips drops mean reward from $-0.10$ to $-0.74$ ($\Delta = 0.64$), so the score captures musically meaningful temporal structure beyond the global power spectrum.

\section{Input Ablation: Full Table}
\label{app:feature_modality}

Table~\ref{tab:feature_modality} reports the seven-variant input ablation that backs the summary in Section~\ref{sec:eval_internal}. All variants share the MLP head template, four-dataset training split, and evaluation protocol. They differ only in the input feature stack.

\begin{table}[h]
\centering
\caption{TuneJury input ablation. CLAP text alone is barely above chance. The six audio-containing variants cluster within a $0.013$ band ($0.695$--$0.708$ Overall). Each row is a single-seed retrain at seed $42$, so absolute accuracies differ from the released checkpoint ($0.7086$ overall, $0.800$ on Music Arena, Section~\ref{sec:eval_internal}) within single-seed noise (${\sim}0.01$ on $n{=}2{,}035$; the $n{=}20$ Music Arena cell carries ${\sim}0.10$ seed variation). A7 is the released architecture. \textbf{Bold} / \underline{underline}: best / 2nd per column, with all tied cells marked.}
\label{tab:feature_modality}
\footnotesize
\begin{tabular}{llccccc}
\toprule
ID & Features & Overall & Music Arena & MusicPrefs & AIME & SongEval \\
\midrule
A1 & CLAP audio only          & $\underline{0.705}$ & $\underline{0.800}$ & $\underline{0.733}$ & $\underline{0.671}$ & $0.888$ \\
A2 & MERT only                & $0.695$ & $0.700$ & $0.650$ & $\underline{0.671}$ & $0.884$ \\
A3 & CLAP text only           & $0.515$ & $0.550$ & $0.544$ & $0.511$ & $0.518$ \\
A4 & CLAP audio $+$ MERT      & $0.701$ & $0.700$ & $0.684$ & $\underline{0.671}$ & $\underline{0.908}$ \\
A5 & CLAP audio $+$ CLAP text & $\mathbf{0.708}$ & $\mathbf{0.850}$ & $\mathbf{0.767}$ & $\underline{0.671}$ & $0.884$ \\
A6 & MERT $+$ CLAP text       & $0.698$ & $\underline{0.800}$ & $0.689$ & $0.667$ & $0.896$ \\
A7 & CLAP audio $+$ MERT $+$ CLAP text & $\underline{0.705}$ & $0.700$ & $0.670$ & $\mathbf{0.674}$ & $\mathbf{0.924}$ \\
\bottomrule
\end{tabular}
\end{table}

\section{External Evaluation: Details}
\label{app:cmi_external_detail}

\paragraph{Disjointness verification.}
Disjointness from our $14{,}346$-audio-file training pool is verified at three levels: file / sample identifier, prompt text (case-insensitive), and byte-level MD5 of the audio. Overlap is zero on each level for PAM, MusicEval, and CMI-Pref test. CMI-RewardBench's $1{,}340$-pair Music Arena split overlaps our training distribution and is handled in the next paragraph.

\paragraph{Music Arena: bench-disjoint training and the leave-MA-out diagnostic.}
All $1{,}340$ of CMI-RewardBench's MA pairs fall into our raw MA pool (same 2025-07 to 2026-01 window, $2{,}039$ live battles). We remove all $1{,}340$ \texttt{battle\_uuid}s from our entire MA pool (train, validation, test) before training every TuneJury variant in Table~\ref{tab:training_mix}, so every MA cell is item-disjoint from the CMI-RewardBench MA test split. Leave-MA-out ($0.6910$) drops every MA pair from training and isolates MA's training contribution. The 2026-02/03 batches ($799$ pairs after \texttt{TIE}\,/\,\texttt{BOTH\_BAD} exclusion) serve as a stricter post-cutoff probe below.

\paragraph{Training-mix design space.}
Table~\ref{tab:training_mix} extends the leave-one-out study to the external CMI-RewardBench axes. No single training mix dominates: leave-(MP$+$MA)-out tops PAM SRCC, leave-SE-out tops MusicEval SRCC and MA pairwise accuracy, and leave-(SE$+$MA)-out tops CMI-Pref. Three of the four leaders exceed CMI-RewardBench leader SongEval-RM on their respective axes. The leave-SongEval-out gains on PAM and MusicEval support the gap-filter distortion discussion (Section~\ref{sec:discussion}). We release the four-dataset checkpoint as the single reward signal that backs every Mode~1--3 demonstration, because it maximizes per-dataset internal coverage and avoids tying the released artifact to a single external axis. A principled resolution to the leave-out trade-off is the mixed-supervision design in Section~\ref{sec:discussion} (Open directions).

\begin{table}[h]
\centering
\caption{Training-mix ablation across the four external
CMI-RewardBench splits. Each TuneJury row is a separate $2$- or $3$-dataset retrain of the same architecture on the listed training subset (omitted datasets are not in training). \emph{Primary deployed} is the released four-dataset checkpoint. \textbf{Bold}/\underline{underline} mark best/$2$nd per column among the leave-one-out and leave-two-out rows (Primary excluded from the ablation ranking).}
\label{tab:training_mix}
\footnotesize
\setlength{\tabcolsep}{4pt}
\begin{tabular}{lcccc}
\toprule
& \multicolumn{2}{c}{Musicality SRCC} & \multicolumn{2}{c}{Pairwise accuracy} \\
\cmidrule(lr){2-3} \cmidrule(lr){4-5}
Training mix & PAM & MusicEval & CMI-Pref & Music Arena \\
\midrule
Primary deployed (MA $+$ MP $+$ AIME $+$ SE) & $0.6100$ & $0.6687$ & $0.7140$ & $0.7194$ \\
Leave-AIME-out (MA $+$ MP $+$ SE) & $0.4808$ & $0.6771$ & $0.7200$ & $\underline{0.7134}$ \\
Leave-MP-out (MA $+$ AIME $+$ SE) & $0.6238$ & $0.6539$ & $0.7180$ & $0.7000$ \\
Leave-MA-out (MP $+$ AIME $+$ SE) & $\underline{0.6381}$ & $\underline{0.7100}$ & $\underline{0.7380}$ & $0.6910$ \\
Leave-SE-out (MA $+$ MP $+$ AIME) & $0.6331$ & $\mathbf{0.7154}$ & $0.7120$ & $\mathbf{0.7149}$ \\
Leave-(SE$+$MA)-out (MP $+$ AIME) & $0.5636$ & $0.6944$ & $\mathbf{0.7480}$ & $0.6791$ \\
Leave-(MP$+$MA)-out (AIME $+$ SE) & $\mathbf{0.6999}$ & $0.6536$ & $0.7100$ & $0.6993$ \\
\bottomrule
\end{tabular}\\
\smallskip
{\footnotesize\raggedright Every TuneJury row is retrained bench-clean (Section~\ref{sec:tunejury}): all $1{,}340$ CMI-RewardBench MA test \texttt{battle\_uuid}s are removed from the MA training pool (verified $0$ overlap), so every Music Arena cell is item-disjoint from that split.}
\end{table}

\paragraph{Pairwise-accuracy view of PAM and MusicEval.}
The PAM and MusicEval columns above report SRCC against per-clip musicality MOS. Because TuneJury is trained with a pairwise objective, we additionally compute \emph{pairwise accuracy} on the same splits by counting clip-pair orderings that agree with the ground-truth MOS ordering (Table~\ref{tab:pw_acc_pam_me}; PAM gives $121{,}016$ pairs and MusicEval $79{,}754$ on the released rows).

\begin{table}[h]
\centering
\caption{Pairwise accuracy on PAM and MusicEval (every
distinct clip pair). TuneJury rows show training-mix variants. CMI-RM included as a closest-comparable baseline. PAM score, Audiobox-Aesthetics, and SongEval-RM are deferred (require re-running their per-clip predictions through CMI-RewardBench's \texttt{inference\_benchmark.py}). MuQ-Eval-A1 per-clip predictions are available in \href{https://github.com/yonghyunk1m/TuneJury/blob/main/applications/baselines/results/muqeval_a1/summary.json}{\texttt{applications/baselines/results/muqeval\_a1/summary.json}}; pairwise-accuracy aggregation into this table is left to future work. \textbf{Bold}/\underline{underline} mark best/$2$nd per column among the item-disjoint rows. $(\mathit{italic})$ marks the in-distribution CMI-RM MusicEval cell (excluded from the ranking).}
\label{tab:pw_acc_pam_me}
\footnotesize
\setlength{\tabcolsep}{6pt}
\begin{tabular}{lcc}
\toprule
Model / training mix & PAM & MusicEval \\
\midrule
CMI-RM (TLRA, $+$110K pseudo, MuQ)      & $\underline{0.7427}$ & $(\mathit{0.8365})$ \\
\midrule
TuneJury, Primary deployed (T$+$A)                & $0.7193$ & $0.7521$ \\
TuneJury, Leave-AIME-out                          & $0.6695$ & $0.7589$ \\
TuneJury, Leave-MP-out                            & $0.7271$ & $0.7469$ \\
TuneJury, Leave-MA-out                            & $0.7327$ & $0.7734$ \\
TuneJury, Leave-SE-out                            & $0.7312$ & $\underline{0.7748}$ \\
TuneJury, Leave-(MP$+$MA)-out (CLAP$+$MERT)       & $\mathbf{0.7577}$ & $0.7472$ \\
TuneJury, MuQ-encoder swap (no-MA)                & $0.7225$ & $\mathbf{0.8126}$ \\
\bottomrule
\end{tabular}
\end{table}

\noindent Pairwise accuracy is the natural reading for a pairwise-trained reward model: the training objective directly optimizes pair ordering. Its values are not comparable to the SRCC columns, since chance sits at $0.5$ for pairwise accuracy and at $0$ for SRCC. Among the item-disjoint rows, MuQ-MuLan-large tops MusicEval ($0.8126$) and Leave-(MP$+$MA)-out tops PAM ($0.7577$), with CMI-RM trailing on PAM ($0.7427$) while its MusicEval cell is in-distribution. Pairwise accuracy for the remaining baselines is left to future work.

\paragraph{Encoder swap probe.}
Holding the head template and training mix fixed, we swap the $2048$-d CLAP$+$MERT stack for MuQ-MuLan-large~\cite{tencent2025muq} (${\sim}663$\,M, $1024$-d joint audio$+$text). The MuQ head uses widths $[512, 256, 128, 64]$ (${\sim}0.7$\,M params) vs.\ $[1024, 512, 256, 128]$ for CLAP$+$MERT (${\sim}2.8$\,M), trained on the same MusicPrefs $+$ AIME $+$ SongEval mix. Single-seed comparison in Table~\ref{tab:encoder_swap}.

\begin{table}[h]
\centering
\caption{Encoder swap probe: holding the head template
($4$ hidden layers with widths scaled to encoder dim) and training mix fixed, replacing LAION-CLAP$+$MERT ($2048$-d input) with MuQ-MuLan-large ($1024$-d joint audio$+$text input). Both rows use the same MusicPrefs $+$ AIME $+$ SongEval $3$-dataset training mix (no Music Arena), so all five cells exclude Music Arena from training (leave-MA-out). The last column is the 2026-02/03 post-cutoff Music Arena slice ($799$ pairs), evaluated under the same Music Arena pairwise protocol. \textbf{Bold} marks the better of the two per column.}
\label{tab:encoder_swap}
\footnotesize
\setlength{\tabcolsep}{3pt}
\renewcommand{\arraystretch}{0.95}
\begin{tabular}{lccccc}
\toprule
& \multicolumn{2}{c}{Musicality SRCC} & \multicolumn{3}{c}{Pairwise accuracy} \\
\cmidrule(lr){2-3} \cmidrule(lr){4-6}
Encoder & PAM & MusicEval & CMI-Pref & CMI-RewardBench MA & MA 2026-02/03 \\
\midrule
LAION-CLAP $+$ MERT ($2048$-d)  & $\mathbf{0.6381}$ & $0.7100$           & $0.7380$           & $0.6910$           & $0.5385$ \\
MuQ-MuLan-large ($1024$-d)      & $0.6146$          & $\mathbf{0.7848}$  & $\mathbf{0.7680}$  & $\mathbf{0.7004}$  & $\mathbf{0.5671}$ \\
\bottomrule
\end{tabular}
\end{table}

\noindent MuQ-MuLan-large matches or beats CLAP$+$MERT on four of five OOD axes (${+}0.075$ MusicEval SRCC, ${+}0.030$ CMI-Pref, ${+}0.009$ CMI-RewardBench MA, ${+}0.029$ post-cutoff MA) at half the input dimension, with a small $-0.024$ PAM SRCC regression. Its MusicEval SRCC ($0.7848$) exceeds the CMI-RewardBench leader SongEval-RM ($0.6949$) by ${+}0.090$. Music-text contrastive pretraining at the ${\sim}663$\,M scale appears to transfer more strongly to OOD naturalistic musicality MOS than CLAP$+$MERT at matched supervision. We release the MuQ-MuLan encoder-swap checkpoint (\texttt{tunejury\_muq\_leave\_MA.pt}) alongside CLAP$+$MERT, the configuration trained on the full four-dataset mix behind every application.

\paragraph{Inference-input scope across splits.}
CMI-RM's TLRA architecture accepts null inputs for missing modalities. PAM and MusicEval provide no lyrics or reference audio in their pairs, so on those benchmarks CMI-RM effectively operates at TuneJury's input scope. CMI-RewardBench Music Arena carries lyrics for ${\sim}55\%$ of pairs (no reference audio), and CMI-Pref test carries lyrics or reference audio for ${\sim}75\%$ of pairs, so CMI-RM retains a partial-to-full TLRA-channel advantage on those two splits that TuneJury does not access by design.

\paragraph{Post-cutoff Music Arena probe.}
As a stricter generalization probe beyond the bench-clean CMI-RewardBench MA split, we additionally collected the 2026-02 and 2026-03 Music Arena batches ($799$ pairs with valid A~vs.~B preference after excluding \texttt{TIE}\,/\,\texttt{BOTH\_BAD} verdicts), a slice whose battles post-date both our feature cache and CMI-RewardBench's training cutoff. Pairwise accuracy on this post-cutoff slice is as follows: released TuneJury (text $+$ audio) reaches $0.5369$, released TuneJury (audio-only, zero-vector empty-prompt protocol of Section~\ref{sec:tunejury}) $0.5307$, CMI-RM~\cite{ma2026cmirewardbench} $0.5614$, leave-MA-out TuneJury $0.5385$ (no MA in training), and the MuQ-MuLan encoder-swap variant $0.5671$ (no MA in training, unchanged from Table~\ref{tab:encoder_swap}, where it also leads on CMI-RewardBench Music Arena). Both TuneJury and CMI-RM drop substantially from the CMI-RewardBench Music Arena ladder (TuneJury $0.7194$, CMI-RM $0.7343$) to the high-$0.5$ regime on this post-cutoff slice. The drop is a difficulty shift driven by newer generators entering the arena after our training cutoff (decomposed in the next paragraph) rather than a TuneJury-specific regression.

\paragraph{Post-cutoff failure decomposition.}
\label{app:postcut_decomposition}
Three diagnostics localize the gap. \emph{(i)~Covariate shift}: four of eleven post-cutoff systems (ACE-Step Turbo Continuous, Lyria~$3$-$30$s, Lyria~$3$ Pro preview, Sonauto~v$3$ preview) are unseen in training, with significant per-system bias ($p<10^{-3}$): TuneJury disagrees with the human vote on $73.6\%$ of pairs in which the human picks against ACE-Step, and on $83.8\%$ of pairs in which the human picks MusicGen-medium. \emph{(ii)~Encoder-space drift}: MusicGen-medium sits $0.18$ cosine units further from the in-distribution CLAP centroid than the median trained system, while MERT cosines are uniform across systems, isolating CLAP as the drifting encoder. \emph{(iii)~Partial margin gradient}: agreement rises with the released TuneJury margin $|\Delta r|$, from $51.3\%$ at $|\Delta r|{<}0.5$ to $62.0\%$ at $|\Delta r|{>}1.2$, a $10.7$\,pp gradient that is informative but well below the $0.7086$ in-distribution test accuracy (Section~\ref{sec:eval_internal}), so margin-based abstention recovers only part of the gap.

\begin{figure}[t]
\centering
\includegraphics[width=\linewidth]{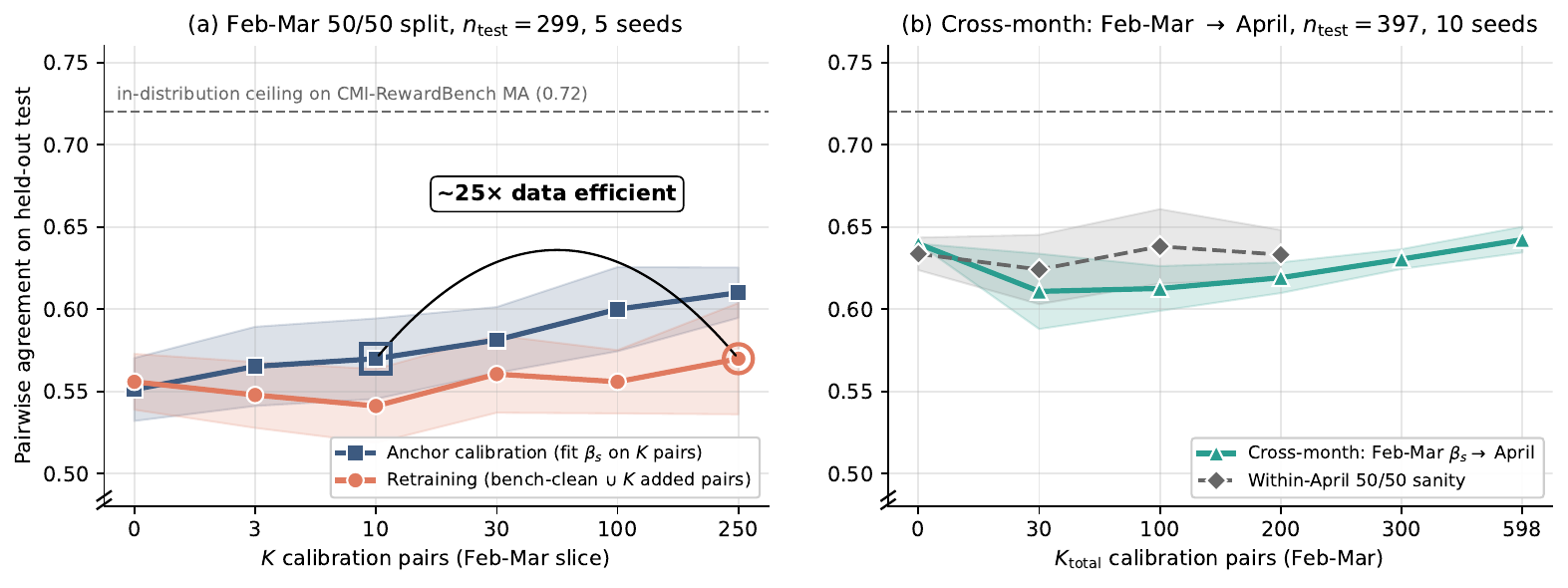}
\caption{OOD recovery on the post-cutoff Music Arena slice. Shaded bands: $95\%$ CI. Dashed reference: released TuneJury's in-distribution agreement on CMI-RewardBench MA ($0.72$). \emph{(a)} Feb--Mar $50/50$ split ($n_\text{test}{=}299$, five seeds): anchor at $K{=}10$ matches retraining at $K{=}250$ (${\sim}25\times$ data-efficiency edge over the swept $K$-grid), anchor at $K{=}30$ exceeds retraining's best swept value. \emph{(b)} Cross-month: Feb--Mar $\beta_s$ fits on April~2026 ($n_\text{test}{=}397$, ten seeds), with a within-April sanity probe overlaid. The released checkpoint already reaches ${\sim}64\%$ raw on April. Feb--Mar fits do not transfer (per-system swings span $-10$ to $+19$\,pp, net cancels).}
\label{fig:ood_scaling}
\end{figure}

\paragraph{Anchor calibration recovers post-cutoff agreement without retraining.}
\label{app:anchor_calibration}
Fitting per-system Bradley--Terry offsets on top of the frozen released TuneJury recovers post-cutoff agreement at substantially better data efficiency than from-scratch retraining (anchor at $K{=}10$ already matches a retrain at $K{=}250$ pairs; Figure~\ref{fig:ood_scaling}, Table~\ref{tab:ood_scaling}). We call this \emph{anchor calibration} because one in-distribution system is held at $\beta{=}0$ for identifiability, anchoring the score scale to its training-time meaning. Post-hoc reward-model calibration has been studied in language models for length bias~\cite{huang2025posthocreward} and for per-policy response bias via continued training on Arena-Elo-derived preferences~\cite{zhu2025charm}. We share the post-hoc framing but target per-system temporal-cutoff bias and fit per-system offsets directly on a small set of post-cutoff calibration pairs, without continued training of the underlying reward head. Motivated by Diagnosis~(i) above, we treat the released TuneJury score as the offset in a Bradley--Terry model~\cite{bradley1952rank} with a per-system bias,
\begin{equation}
P(a \succ b) =
\sigma\!\big( (r(a) - \beta_{s_a}) - (r(b) - \beta_{s_b}) \big),
\end{equation}
and fit $\{\beta_s\}$ by L-BFGS~\cite{liu1989lbfgs} on $K$ post-cutoff calibration pairs ($\ell_2$ regularization $\lambda{=}1.0$, $<1$\,s CPU; encoders frozen). Of the $799$ decisive Feb--Mar pairs (Table~\ref{tab:encoder_swap}), $598$ involve at least one in-distribution anchor system and form the anchor-calibration pool used in the rest of this section. Figure~\ref{fig:ood_scaling} compares (R)~retraining from scratch on bench-clean ($571$) $\cup$ $K$ added pairs against (A)~anchor calibration on the released TuneJury, both on a $50/50$ held-out split of this $598$-pair slice ($n_\text{test}{=}299$, five seeds). Anchor calibration recovers ${\sim}3$\,pp at $K{=}30$ and ${\sim}5$\,pp at $K{=}100$ (Table~\ref{tab:ood_scaling}). Anchor at $K{=}10$ ($57.0$) already matches retraining's best swept $K{=}250$ ($57.0$), so within the swept $K \in \{0, 3, 10, 30, 100, 250\}$ grid this is a ${\sim}25\times$ data-efficiency edge. The $K{=}250$ cap reflects the $299$-pair calibration half of the split, so further retraining gains at $K>250$ cannot be ruled out.

\begin{table}[h]
\centering
\caption{Post-cutoff agreement under the two recovery strategies (mean $\pm$ std
over $5$ seeds; held-out $n_\text{test}=299$). Both rows reproduced against
the released checkpoint \texttt{tunejury.pt} (md5 \texttt{0524e60}) using
\href{https://github.com/yonghyunk1m/TuneJury/tree/main/applications/anchor_calibration}{\texttt{applications/anchor\_calibration/}} (\href{https://github.com/yonghyunk1m/TuneJury/blob/main/applications/anchor_calibration/run_experiment.py}{\texttt{run\_experiment.py}} for A and
\href{https://github.com/yonghyunk1m/TuneJury/blob/main/applications/anchor_calibration/retrain_ksweep.py}{\texttt{retrain\_ksweep.py}} for R). Anchor calibration beats retraining at every $K{\geq}3$, and at $K{=}30$ anchor already exceeds retraining's best swept value at $K{=}250$. \textbf{Bold} marks the better row per column.}
\label{tab:ood_scaling}
\footnotesize
\setlength{\tabcolsep}{4pt}
\begin{tabular}{lcccccc}
\toprule
$K$ & $0$ & $3$ & $10$ & $30$ & $100$ & $250$ \\
\midrule
Retraining (R) & $\mathbf{55.6\pm1.7}$ & $54.8\pm2.0$ & $54.1\pm2.3$ & $56.1\pm2.4$ & $55.6\pm2.0$ & $57.0\pm3.5$ \\
Anchor calib. (A) & $55.1\pm2.0$ & $\mathbf{56.5\pm2.5}$ & $\mathbf{57.0\pm2.5}$ & $\mathbf{58.1\pm2.0}$ & $\mathbf{60.0\pm2.6}$ & $\mathbf{61.0\pm1.6}$ \\
\midrule
Gap ($A - R$) & $-0.5$ & $+1.8$ & $+3.5$ & $+1.6$ & $+4.5$ & $+4.2$ \\
\bottomrule
\end{tabular}
\end{table}

\paragraph{Recommended protocol under continual generator drift.}
Use the released TuneJury unchanged for in-distribution scoring, and refit $\beta_s$ on ${\sim}100$ post-cutoff calibration pairs against an in-distribution anchor system (e.g., Sonauto~v$2$ held at $\beta{=}0$). A smaller refit at ${\sim}30$ pairs already exceeds a from-scratch retrain on $250$ post-cutoff pairs (Table~\ref{tab:ood_scaling}). Mode~1--3 results in Section~\ref{sec:applications} evaluate in-distribution selection / optimization and are unaffected.

\paragraph{Cross-month generalization to truly held-out April~2026.}
The Feb--Mar $50/50$ split shares a month, so a strict reading is that anchor calibration may memorize within-month structure. We reuse the \emph{same} Feb--Mar $\beta_s$ fits on a held-out month, the April~2026 Music Arena release (Hugging Face \texttt{music-arena/music-arena-dataset} config \texttt{2026\_04}; $397$ decisive pairs after excluding \texttt{TIE}\,/\,\texttt{BOTH\_BAD} and audio-withheld battles). One April system (\texttt{sao}, the Stable Audio Open battle tag) is unseen in both training and Feb--Mar, so it gets $\beta{=}0$.

Released TuneJury reaches $0.6398$ raw on April, already much higher than its $0.551$ raw on the Feb--Mar $598$-pair anchor-calibration pool (Table~\ref{tab:ood_scaling}, anchor at $K{=}0$), indicating that the OOD difficulty is concentrated on the Feb--Mar slice (which contains the newest generators at the time) rather than uniformly distributed across post-cutoff months. Feb--Mar anchor calibration at $K_{\text{total}}{=}30$ moves April to $0.6108\pm0.0350$, a ${\sim}2.9$\,pp \emph{regression} on April. The cross-month curve recovers to $0.6423\pm0.0120$ only at $K_{\text{total}}{=}598$, essentially matching the raw baseline (Table~\ref{tab:cross_month}). A within-April sanity probe ($50/50$, $n_{\text{test}}{=}199$) is similarly flat across $K$ ($0.634$ at $K{=}0$, $0.633$ at $K{=}200$). Anchor calibration does not transfer from Feb--Mar to April at small $K$: per-system biases fit on Feb--Mar generators do not improve April agreement, and the released checkpoint is already close to its in-month agreement ceiling on April without calibration.

\begin{table}[h]
\centering
\caption{Cross-month application of Feb--Mar anchor-calibration fits
(mean $\pm$ std, ten seeds; regenerated against the released
checkpoint). \emph{Cross-month}: fit $\beta_s$ on $K_{\text{total}}$
Feb--Mar pairs, evaluate on all $397$ April decisive pairs.
\emph{Within-April}: $50/50$ sanity probe on April
($n_{\text{test}}{=}199$). On April the released raw is already
${\sim}64\%$, and anchor calibration at small $K$ regresses below raw
before recovering to baseline by $K{=}598$. Per-system biases
calibrated on Feb--Mar do not transfer.}
\label{tab:cross_month}
\footnotesize
\setlength{\tabcolsep}{4pt}
\begin{tabular}{lccccc}
\toprule
$K_{\text{total}}$ & $0$ & $30$ & $100$ & $200$ & $598$ \\
\midrule
Cross-month (Feb--Mar $\to$ April) & $64.0$       & $61.1\pm3.5$ & $61.3\pm2.1$ & $61.9\pm1.4$ & $\mathbf{64.2\pm1.2}$ \\
Within-April ($50/50$)            & $63.4\pm1.5$ & $62.4\pm3.2$ & $63.8\pm3.5$ & $63.3\pm2.3$ & -- \\
\bottomrule
\end{tabular}
\end{table}

\paragraph{Post-cutoff battles are intrinsically harder.}
The cross-month plateau ($K{=}598$ at $0.642$, Figure~\ref{fig:ood_scaling}(b)) sits ${\sim}8$\,pp below the $0.72$ in-distribution agreement on CMI-RewardBench MA, but the gap is not pure model failure: post-cutoff battles are harder for both TuneJury and the human voters. (i)~Released TuneJury's $|\Delta r|$ compresses on the post-cutoff slices (Table~\ref{tab:intrinsic_difficulty}: mean drops from $1.148$ in-distribution to $0.647$ on Feb--Mar and $0.721$ on April; the share of $|\Delta r|{<}1.0$ rises from $62\%$ in-distribution to $80\%$ on Feb--Mar and $72\%$ on April). The heavier compression on Feb--Mar mirrors its lower raw agreement (paragraph above): the model is least confident exactly where it is least accurate. (ii)~Human voters disagree more: of $674$ raw April battles, $35\%$ are non-decisive ($25.4\%$ \texttt{BOTH\_BAD} $+$ $9.9\%$ \texttt{TIE}). The April generator population has likely converged into a tighter perceptual quality band, so the $0.72$ CMI-RewardBench MA reference is a pre-cutoff label-noise ceiling rather than an April-specific one.

\begin{table}[h]
\centering
\caption{TuneJury margin distribution shifts toward the boundary on
post-cutoff battles, consistent with intrinsic task difficulty rather
than a pure TuneJury failure mode. In-distribution baseline is the full
non-tie $n{=}2{,}035$ four-dataset held-out test (Section~\ref{sec:tunejury}).
$|\Delta r|$ is the released TuneJury's absolute pairwise margin. \textbf{Bold}: the most boundary-shifted value per column.}
\label{tab:intrinsic_difficulty}
\footnotesize
\begin{tabular}{lccccc}
\toprule
Test set & $n$ & mean $|\Delta r|$ & median $|\Delta r|$ & $|\Delta r|{<}0.5$ & $|\Delta r|{<}1.0$ \\
\midrule
In-distribution test (four-dataset) & $2{,}035$ & $1.148$ & $0.728$ & $36\%$ & $62\%$ \\
Feb--Mar pool               & $598$ & $\mathbf{0.647}$ & $\mathbf{0.496}$ & $\mathbf{51\%}$ & $\mathbf{80\%}$ \\
April~2026 (cross-month)     & $397$ & $0.721$ & $0.564$ & $44\%$ & $72\%$ \\
\bottomrule
\end{tabular}
\end{table}

The residual gap to $0.72$ is therefore a mix of per-system bias and encoder drift on the TuneJury side (Diagnoses~i--ii above) and a falling April-specific human ceiling. Anchor calibration addresses the per-system bias, the encoder-swap variants target the drift, and the human ceiling is a property of the test distribution, not the reward model.

\paragraph{No catastrophic forgetting in a bench-clean fold-in probe.}
Retraining a TuneJury-style head on the bench-clean MA train split ($571$ pairs) augmented with the $995$ post-cutoff pairs (three seeds) moves Feb--Mar $+5$\,pp and April $+7$\,pp while leaving the bench-clean MA test split within noise ($n{=}20$). The probe is MA-local, so it bounds forgetting for the Music Arena component rather than the full four-dataset mix.

\paragraph{Why anchor calibration outperforms naive retraining.}
The failure mode is encoder-distribution drift on top of an additive per-system bias, not missing data. Adding $K{=}250$ post-cutoff pairs to the $571$-pair bench-clean training pool moves retrain accuracy only $+1.4$\,pp from the $K{=}0$ baseline (Table~\ref{tab:ood_scaling}, R~row), while a per-system bias term captures most of the recoverable signal at $K{=}30$ (anchor $58.1$ exceeds the retrain $K{=}250$ ceiling $57.0$). Encoder drift itself is better addressed by the encoder-swap variants in Table~\ref{tab:encoder_swap} (MuQ-MuLan-large reaches $0.5671$ on the same slice, an encoder-swap probe).

\paragraph{Prompt format examples.}
One example per external split illustrates why arena-style prompts are training-aligned and the others are OOD. \emph{CMI-Pref test} (free-form arena-style request): \texttt{``melodic japanese folk synth-pop''}. \emph{MusicEval} (stylistic spec): \texttt{``A lively, short summer piano solo piece, ideal for indoor performance''}. \emph{PAM} (post-hoc caption of existing audio): \texttt{``A digital drum is playing a simple rhythm along with a synth bassline.''} MusicEval and PAM both differ systematically from the arena-style requests in our training mix, consistent with the text-branch effect in Section~\ref{sec:eval_external}.

\paragraph{Text-input dropout retrain.}
A TuneJury variant trained with $30\%$ text-input dropout underperforms the released checkpoint on every external split under both prompt protocols, and the SRCC gap between the with-prompt and empty-prompt protocols widens on PAM ($0.063 \to 0.122$) rather than narrowing. We report this as a negative result and keep the no-dropout variant.

\paragraph{AIME held-out sanity check: per-baseline and per-axis breakdown.}
\label{app:aime_baselines}
On the $1{,}560$-pair AIME~\cite{grotschla2025aime} held-out test, released TuneJury (T$+$A) reaches $0.6744$, surpassing every baseline by $2.2$ to $6.4$\,pp. The baselines span $0.6103$--$0.6526$ under their published protocols, with SongEval-RM Musicality the strongest at $0.6526$ (per-axis breakdown in Table~\ref{tab:aime_per_axis}). AIME is in-distribution for TuneJury and out-of-distribution for the baselines (AIME is not in any baseline's training data), so this is a sanity check rather than a head-to-head claim. The leave-AIME-out retrain, OOD like the baselines, reaches $0.625$ on the same split, inside the baseline span (Table~\ref{tab:internal_per_dataset}). CMI-RM runs with null lyrics and reference-audio embeddings, matching its inference setup on PAM and MusicEval (CMI-RewardBench Music Arena carries lyrics for ${\sim}55\%$ of pairs, which CMI-RM does encode). Audiobox-Aesthetics reports four axes (CE, CU, PC, PQ), with PC uncorrelated with preference ($0.5000$) and CE/CU/PQ clustering in $0.60$--$0.62$. SongEval-RM's five axes cluster in $0.64$--$0.65$ with Musicality the per-axis best. CMI-RM reports alignment and musicality, with the alignment axis trailing by ${\sim}3$\,pp ($0.6032$). Audio is cropped to $30$\,s (affects ${\sim}31\%$ of clips, $24$\,GB GPU memory cap on long human references). Baselines: \href{https://huggingface.co/microsoft/msclap}{\texttt{microsoft/msclap}}, \href{https://huggingface.co/facebook/audiobox-aesthetics}{\texttt{facebook/audiobox-aesthetics}}, \href{https://huggingface.co/OpenMuQ/MuQ-large-msd-iter}{\texttt{OpenMuQ/MuQ-large-msd-iter}}, and \href{https://huggingface.co/HaiwenXia/CMI-RM}{\texttt{HaiwenXia/CMI-RM}}.

\begin{table}[h]
\centering
\caption{Per-axis pairwise accuracy on AIME held-out test ($1{,}560$ pairs). The bolded entry per baseline is the preference-aligned headline axis used in the text above. TuneJury (T$+$A) is in-distribution for AIME (AIME is in our training mix) and therefore excluded from this OOD baseline table.}
\label{tab:aime_per_axis}
\footnotesize
\setlength{\tabcolsep}{6pt}
\begin{tabular}{llc}
\toprule
Baseline & Axis & Pairwise accuracy \\
\midrule
PAM score              & zero-shot                  & $0.6442$ \\
\midrule
Audiobox-Aesthetics    & CE (Content Enjoyment)     & $\mathbf{0.6103}$ \\
                       & CU (Content Usefulness)    & $0.6192$ \\
                       & PC (Production Complexity) & $0.5000$ \\
                       & PQ (Production Quality)    & $0.6000$ \\
\midrule
SongEval-RM            & Coherence                  & $0.6417$ \\
                       & Musicality                 & $\mathbf{0.6526}$ \\
                       & Memorability               & $0.6429$ \\
                       & Clarity                    & $0.6487$ \\
                       & Naturalness                & $0.6442$ \\
\midrule
CMI-RM                 & Alignment                  & $0.6032$ \\
                       & Musicality                 & $\mathbf{0.6333}$ \\
\bottomrule
\end{tabular}
\end{table}

\begin{figure}[t]
\centering
\includegraphics[width=\textwidth]{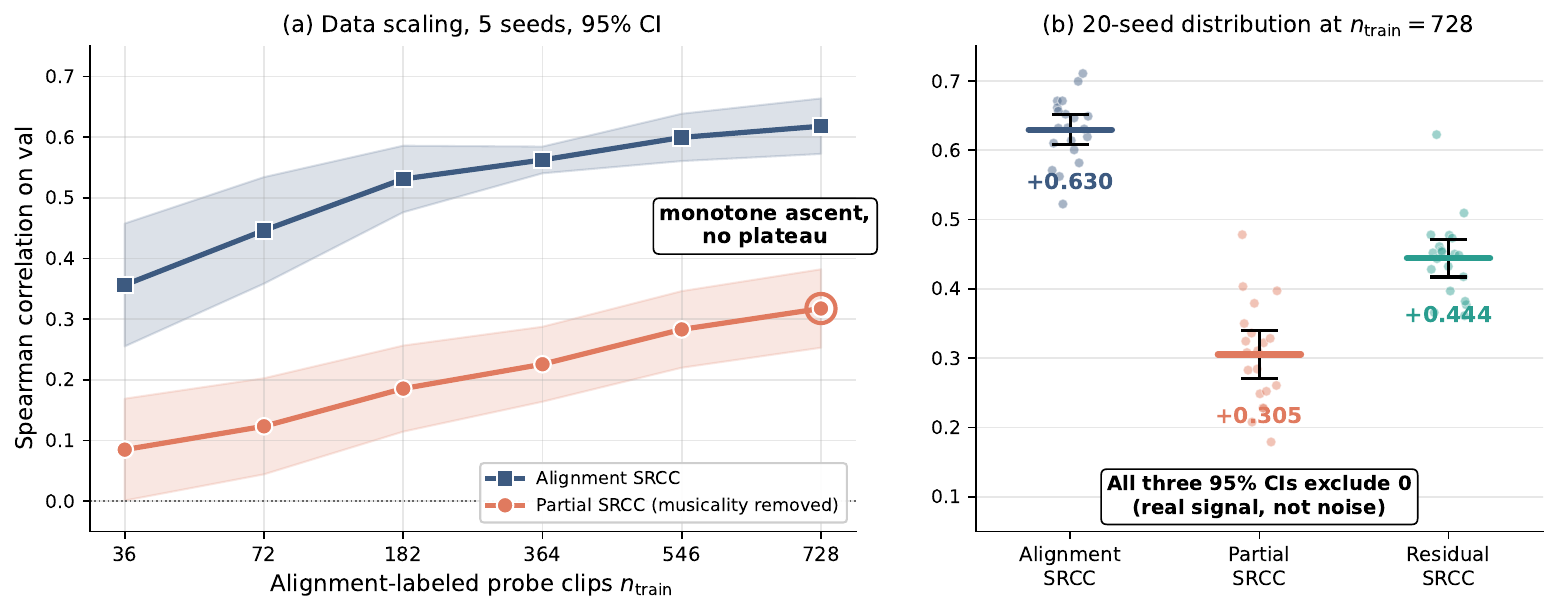}
\caption{Decomposition probe (Appendix~\ref{app:decomp_probe}, body summary in Section~\ref{sec:tunejury}). \emph{(a)}~Data-scaling curve over the alignment-labeled probe set (PAM~$+$~MusicEval, the per-axis MOS pool used only by the probe and distinct from TuneJury's ${\sim}17.5$\,K training pairs). Alignment SRCC (blue) and partial SRCC controlling for musicality (red) both rise monotonically within the available range ($n{=}36$ to $n{=}728$) with no plateau visible at the upper limit, though the probe does not characterize what happens at $10{\times}$ or $100{\times}$ scale. \emph{(b)}~Multi-seed distribution at full data ($n{=}728$ train, $20$-seed). Both partial ($95\%$ CI $[0.271, 0.340]$) and residual ($95\%$ CI $[0.417, 0.472]$) intervals exclude zero.}
\label{fig:decomp_probe}
\end{figure}

\section{Decomposition Probe: Full Details}
\label{app:decomp_probe}

The candidate decomposition splits the score into two parts: an audio-only score (TuneJury with empty prompt) and the text branch's contribution (composite minus audio-only). The probe requires per-clip text-music alignment MOS, which TuneJury's ${\sim}17.5$\,K-pair preference training pool does not provide. The arena-style sources (Music Arena, AIME, MusicPrefs) report a single composite winner per pair with no per-axis decomposition, and SongEval's $5$-axis aesthetic MOS does not include text-music alignment as an axis. CMI-RewardBench's PAM ($n{=}500$) and MusicEval ($n{=}413$) splits are the only pool we had access to with per-clip text-music alignment MOS, totaling $913$ clips. We probe this decomposition on the pool in four stages (Figure~\ref{fig:decomp_probe}).

\paragraph{Stage 1: Post-hoc.}
The audio-only score exceeds the composite on PAM musicality SRCC ($0.6731$ vs.\ $0.6100$; Table~\ref{tab:head_to_head}), but the text branch's contribution (composite minus audio-only) does not recover the alignment axis: SRCC against PAM alignment MOS is ${-}0.30$ and against MusicEval alignment MOS is ${+}0.02$ (deterministic scoring on the full splits).

\paragraph{Stage 2: Cross-distribution supervised.}
Training a fresh MLP head on alignment MOS from one of \{PAM, MusicEval\} and testing on the other (single training run, seed $42$) does not transfer between splits: SRCC is ${+}0.18$ for PAM$\rightarrow$MusicEval and ${-}0.41$ for MusicEval$\rightarrow$PAM.

\paragraph{Stage 3: Stratified combined ($n{=}913$, $80/20$).}
A supervised head reaches alignment SRCC $0.630$, but its partial Spearman controlling for musicality is only $0.305$: the alignment and musicality MOS are themselves Spearman-correlated at ${+}0.716$, so much of the head's signal is general quality. To isolate the alignment-specific signal, we train the head to predict the alignment residual: alignment MOS minus its linear fit on musicality MOS (slope $\beta{=}0.667$ on the combined pool). This head reaches SRCC $0.444$ on the held-out residual, consistent with an alignment-specific signal in the features that remains after removing the musicality-correlated part. (All three values are $20$-seed means, with $95\%$ CIs shown in Figure~\ref{fig:decomp_probe}(b).)

\paragraph{Stage 4: Data scaling within the probe pool.}
The probe head's partial SRCC rises monotonically from $0.085$ at $n{=}36$ to $0.318$ at $n{=}728$, the upper limit set by the $80/20$ split over the ${\sim}900$-clip alignment-labeled MOS pool (Figure~\ref{fig:decomp_probe}(a), $5$-seed mean). The curve has not yet plateaued at the upper limit, so we do not know where scaling stops. This is a statement about the alignment-supervised probe head only, not about TuneJury's main training, which uses the ${\sim}17.5$\,K-pair preference pool (about $20{\times}$ larger and on a different supervision signal). Multi-head supervision and pseudo-label augmentation are candidate paths to extend the probe (Section~\ref{sec:discussion}, Open directions).

\begin{figure}[h]
\centering
\includegraphics[width=\linewidth]{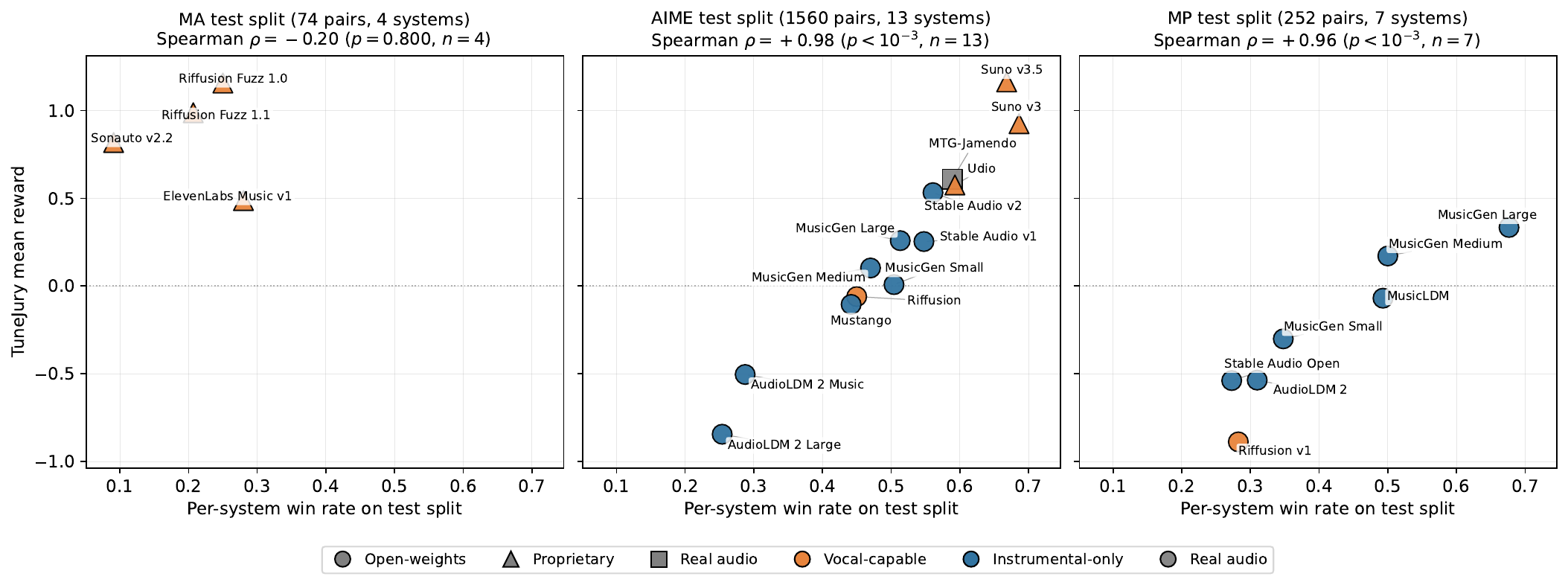}
\caption{Per-system TuneJury reward vs.\ win rate on bench-clean held-out splits. Marker shape: circle $=$ open-weights, triangle $=$ proprietary, square $=$ real audio. Color: orange $=$ vocal-capable, blue $=$ instrumental-only, gray $=$ real audio. AIME ($\rho{=}{+}0.98$, $n{=}13$) and MusicPrefs ($\rho{=}{+}0.96$, $n{=}7$) show high system-rank agreement between TuneJury and human votes at this small system count. Music Arena ($n{=}74$) is too small for a reliable per-system signal.}
\label{fig:per_system_test}
\end{figure}

\section{Per-System Reward Ranking on Held-Out Test Splits}
\label{app:per_system}

We test the Section~\ref{sec:discussion} ``capability proxy'' claim with a per-system reward ranking probe on the three datasets with model labels (Music Arena, AIME, MusicPrefs; SongEval has anonymized labels). For each test pair we score both clips with the released TuneJury, aggregate by source system, and compute Spearman rank correlation against the per-system win rate on the same test split.

On AIME's held-out test split ($1{,}560$ pairs of $15{,}600$ total, $13$ systems with $\ge 200$ comparisons), $\rho = +0.978$ (Pearson $r = +0.97$), with top-$2$ (Suno~v$3.5$, Suno~v$3$) and bottom-$2$ (AudioLDM2-music, AudioLDM2-large) recovered exactly. Both humans and TuneJury place the top-$2$ Suno checkpoints \emph{above} the MTG-Jamendo real-audio baseline ($67$--$69\%$ vs.\ $59\%$ win), consistent with AIME's `real' baseline being predominantly amateur CC audio. On MusicPrefs's held-out test split ($252$ pairs of $2{,}515$ total, $7$ systems), $\rho = +0.964$ (Pearson $r = +0.93$). AIME and MusicPrefs are in-distribution at the dataset level, so this is an internal-consistency check rather than a generalization claim. The Music Arena bench-clean test ($n{=}74$, $4$ systems with non-zero held-out win rate after CMI-RewardBench overlap removal) is too small for a reliable per-system signal. Figure~\ref{fig:per_system_test} shows the per-system scatter for AIME and MusicPrefs.

\paragraph{Lyrics-presence text-proxy probe.}
On the full Music Arena pool ($3{,}060$ pairs, $6{,}120$ clips), grouping by whether the source pair carried a non-empty lyrics field, vocal-requested clips score mean $+0.977$ vs.\ instrumental $+0.536$, a $+0.441$ gap (Welch $t{=}+24.2$). We read this as TuneJury responding to a lyrics-presence textual proxy in the training data, not as evidence of vocal-quality evaluation: vocal-capable generators (Suno, Udio) dominate the lyrics-present pairs in Music Arena, so the gap is consistent with a system-level preference confound rather than per-clip vocal-skill discrimination. The external validation below probes whether any vocal-specific signal exists beyond this text-proxy effect.

\begin{figure}[h]
\centering
\includegraphics[width=\linewidth]{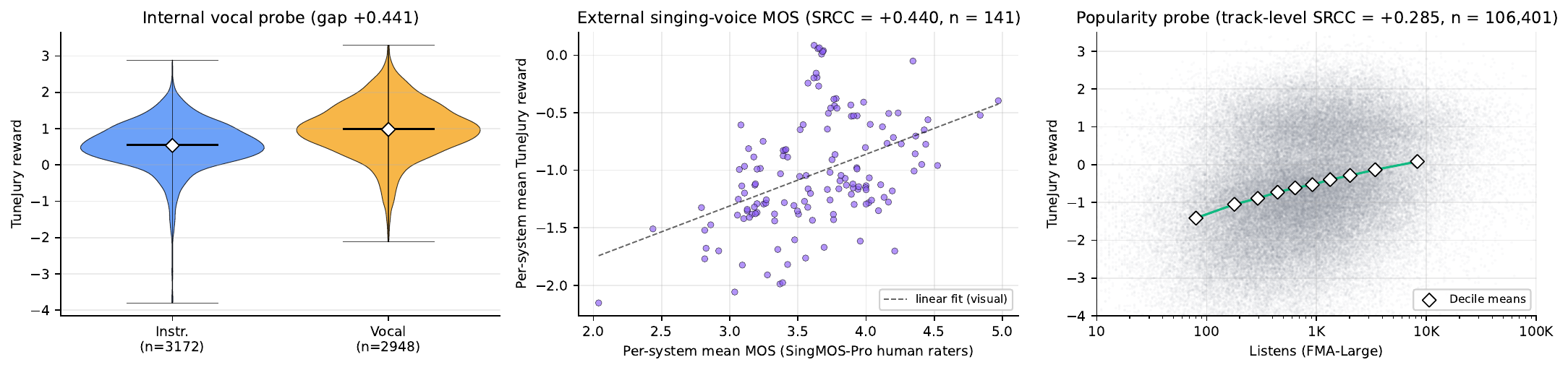}
\caption{Three population-level probes of the TuneJury reward signal (Section~\ref{sec:discussion}).
\emph{Left}: lyrics-presence text-proxy probe. Music Arena clips grouped by whether the source pair's
\texttt{lyrics} field was non-empty (a prompt-level proxy for
vocal-request intent, not a per-clip vocal annotation),
with a gap of ${+}0.441$ reward units at $n{=}6{,}120$. Consistent with a system-level preference confound rather than vocal-quality reward.
\emph{Middle}: external generalization to singing-voice MOS at the system-aggregation level on SingMOS-Pro~\cite{tang2025singmospro} ($n{=}141$ Chinese/Japanese vocal-generation systems), SRCC ${=}\,{+}0.44$. Not direct evidence of vocal-specific features; could reflect general production-quality preference.
\emph{Right}: popularity probe. FMA-Large by listens decile (bottom decile ${-}1.413$ to
top ${+}0.084$, $n{=}106{,}401$). All three probes are population-level.
Per-clip and vocal-specific feature claims are not supported.}
\label{fig:probes}
\end{figure}

\paragraph{External singing-voice MOS validation.}\label{par:external_vocal_validation}
Two external benchmarks probe the TuneJury vocal signal beyond our training distribution. On SingMOS-Pro~\cite{tang2025singmospro} ($n{=}7{,}981$ singing utterances with multi-rater MOS, $141$ singing-voice generation systems across singing voice synthesis, resynthesis, conversion, and ground-truth baselines, Chinese and Japanese), TuneJury per-utterance Spearman is $+0.19$ and per-system Spearman is $+0.44$ (Figure~\ref{fig:probes}, middle panel), both statistically significant ($p < 10^{-3}$). On SVCC~$2025$~\cite{violeta2025svcc25} ($n{=}48$ real-human recordings, $2$ singers $\times$ $6$ vocal techniques), mean TuneJury reward varies across the six techniques (Mixed Voice highest at $-0.11$, Pharyngeal lowest at $-0.70$, ANOVA $F{=}3.82$, $p < 0.01$). TuneJury was not trained on either benchmark. The two probes provide independently obtained population-level signals correlating with human vocal evaluation: SingMOS-Pro shows system-ranking signal, and SVCC-$2025$ shows across-technique discrimination on real human recordings. We do not claim TuneJury isolates vocal-specific quality features: the SingMOS per-system SRCC (${+}0.44$, below dedicated vocal MOS predictors that typically reach $0.6$--$0.8$) could equally reflect general production-quality preferences shared across vocal and instrumental music, and our SongEval training pairs (derived via a mean-gap filter across $5$ aesthetic axes) may already encode indirect vocal-quality signal that propagates into the score. Within the tested scope, TuneJury produces a population-level ranking signal that correlates with human vocal MOS on Chinese/Japanese vocal-generation benchmarks at moderate strength, suitable as a candidate auxiliary signal for system-aggregation comparisons of vocal-generation systems in similar contexts. It is \emph{not} validated for per-clip vocal MOS regression (per-utterance SRCC ${+}0.19$), vocal-specific feature interpretations, generalization to other languages, or as a replacement for dedicated vocal MOS predictors.

\paragraph{Popularity-stratified probe (FMA-Large listens).}
Bucketing the $106{,}401$ released FMA-Large reward scores by \texttt{original\_listens} decile gives a monotone reward gradient with a ${\sim}1.50$-unit gap between the bottom decile ($-1.413$) and the top decile ($+0.084$) (Figure~\ref{fig:probes}, right). The full-distribution Spearman is only $+0.285$, so we read the decile gap rather than the linear correlation. Like the vocal probe, this is population-level and does not validate per-track amateur vs.\ professional discrimination.

\section{Mode 1 Best-of-\texorpdfstring{$N$}{N}: Full Sweep and Extended Analysis}
\label{app:bon_full}

\label{app:mode1_extended}
Table~\ref{tab:apps_mode1_bon_full} reports the full Mode~1 $N\in\{1,2,4,8,16,32\}$ sweep on all four backbones with FAD-CLAP, CLAP score, FAD-MERT, MAD, and TuneJury reward stacked side by side. This is the canonical reference for the main-text Figure~\ref{fig:mode1_bon_sweep} and Section~\ref{sec:apps_bon} numbers. The per-backbone trend across $N$ is discussed in the analysis paragraphs that follow.

\begin{table}[h]
\centering
\caption{Full Mode~1 best-of-$N$ sweep on all four frozen open-weights backbones with the released bench-clean TuneJury as the selector. \textbf{Bold}: best $N$ per metric per backbone. The three distributional metrics (FAD-CLAP, FAD-MERT, and MAD~\cite{huang2025musicprefs}) are computed against SDD-$706$, lower meaning closer. Reward is strictly monotone in $N$ on every backbone, bold at $N{=}32$ throughout.}
\label{tab:apps_mode1_bon_full}
\footnotesize
\setlength{\tabcolsep}{4pt}
\begin{tabular}{lcccccc}
\toprule
Backbone (size, family) & $N$ & FAD-CLAP$\downarrow$ & CLAP score$\uparrow$ & FAD-MERT$\downarrow$ & MAD$\downarrow$ & Reward$\uparrow$ \\
\midrule
\multicolumn{7}{l}{\emph{MusicGen-medium ($1.5$\,B, autoregressive transformer)}} \\
& $1$  & $0.411$ & $0.380$ & $\mathbf{3.88}$ & $1.347$ & $+0.314$ \\
& $2$  & $0.414$ & $0.380$ & $4.14$ & $1.595$ & $+0.570$ \\
& $4$  & $0.397$ & $0.393$ & $4.27$ & $1.973$ & $+0.821$ \\
& $8$  & $0.385$ & $0.397$ & $4.31$ & $1.570$ & $+0.999$ \\
& $16$ & $0.377$ & $0.396$ & $4.86$ & $1.269$ & $+1.188$ \\
& $32$ & $\mathbf{0.372}$ & $\mathbf{0.404}$ & $4.60$ & $\mathbf{1.217}$ & $\mathbf{+1.332}$ \\
\midrule
\multicolumn{7}{l}{\emph{MusicGen-large ($3.3$\,B, autoregressive transformer)}} \\
& $1$  & $0.382$ & $0.386$ & $\mathbf{3.91}$ & $0.772$ & $+0.385$ \\
& $2$  & $0.397$ & $0.390$ & $4.24$ & $1.245$ & $+0.719$ \\
& $4$  & $\mathbf{0.373}$ & $0.397$ & $4.36$ & $1.298$ & $+0.840$ \\
& $8$  & $0.383$ & $\mathbf{0.400}$ & $4.56$ & $0.983$ & $+1.053$ \\
& $16$ & $0.383$ & $0.392$ & $4.74$ & $0.742$ & $+1.250$ \\
& $32$ & $0.381$ & $0.392$ & $4.74$ & $\mathbf{0.717}$ & $\mathbf{+1.370}$ \\
\midrule
\multicolumn{7}{l}{\emph{AudioLDM2-music ($1.1$\,B, latent diffusion)}} \\
& $1$  & $0.755$ & $0.264$ & $7.56$ & $2.283$ & $-0.815$ \\
& $2$  & $0.806$ & $0.276$ & $5.60$ & $2.865$ & $-0.432$ \\
& $4$  & $0.777$ & $0.283$ & $4.89$ & $2.465$ & $-0.050$ \\
& $8$  & $0.685$ & $\mathbf{0.307}$ & $\mathbf{4.73}$ & $\mathbf{0.837}$ & $+0.241$ \\
& $16$ & $0.674$ & $0.300$ & $4.73$ & $0.875$ & $+0.365$ \\
& $32$ & $\mathbf{0.673}$ & $0.297$ & $4.99$ & $1.252$ & $\mathbf{+0.425}$ \\
\midrule
\multicolumn{7}{l}{\emph{ACE-Step Turbo Continuous ($2.4$\,B, continuous-latent DiT)}} \\
& $1$  & $0.725$ & $0.139$ & $6.36$ & $4.962$ & $-0.751$ \\
& $2$  & $0.708$ & $0.163$ & $6.13$ & $3.730$ & $+0.080$ \\
& $4$  & $0.693$ & $0.196$ & $5.85$ & $3.442$ & $+0.608$ \\
& $8$  & $0.606$ & $0.213$ & $4.87$ & $3.963$ & $+0.851$ \\
& $16$ & $0.585$ & $\mathbf{0.214}$ & $4.17$ & $\mathbf{2.364}$ & $+1.064$ \\
& $32$ & $\mathbf{0.581}$ & $0.204$ & $\mathbf{3.88}$ & $2.830$ & $\mathbf{+1.206}$ \\
\bottomrule
\end{tabular}
\end{table}

\paragraph{CLAP score rises with $N$ through $N{=}8$ on every backbone.}
The CLAP score is non-decreasing through $N{=}8$ on every one of the four backbones, after which each backbone peaks at $N{=}8$, $N{=}16$, or $N{=}32$ with small fluctuations ($\le 0.010$). Selecting for TuneJury thus biases samples toward better text alignment as a byproduct of musicality ranking through $N{=}8$ despite TuneJury having no explicit text-alignment training objective.

\paragraph{Distributional fit: three metrics, three patterns.}
The two FAD distances against SDD-$706$ point in opposite directions on most backbones: FAD-CLAP improves through $N{=}32$ on three of four backbones while FAD-MERT moves oppositely on the two MusicGen variants. MAD~\cite{huang2025musicprefs} on MERT embeddings is the only metric where all four backbones move closer to SDD-$706$ net of $N{=}1$ through $N{=}32$, but two of four reach their minimum before $N{=}32$ (AudioLDM2-music at $N{=}8$, ACE-Step Turbo Continuous at $N{=}16$). The cross-encoder interpretation and practitioner recommendation are in Section~\ref{sec:apps_bon}.

\paragraph{Scale generalizes within family at low $N$, with cross-overs at high $N$.}
At every $N \le 8$, MusicGen-large outperforms MusicGen-medium on FAD-CLAP, CLAP score, and Reward. At $N \ge 16$, medium overtakes large on FAD-CLAP and CLAP score while large retains the Reward lead throughout. The selector is identical across scales, so the across-scale shifts reflect each model's $N$-sweep candidate distribution. TuneJury does not need to be retuned across the MusicGen $1.5$--$3.3$\,B range.

\paragraph{Per-doubling gain breakdown.}
We extend each backbone's canonical $N{=}16$ run (seeds $42$--$57$) with $16$ additional candidates at seeds $58$--$73$ to reach $N{=}32$. The intermediate $\Delta_{N{:}8\to16}$ band is $[{+}0.124,{+}0.213]$ (per-backbone values follow from Table~\ref{tab:apps_mode1_bon_full}).

\paragraph{Scope: instrumental-only across the four Mode 1 backbones.}
All four backbones evaluated here generate instrumental music under the prompt-prefix and empty-lyric protocol of Section~\ref{sec:applications}. TuneJury's training mix (Section~\ref{sec:tunejury}) is itself heterogeneous. Music Arena pairs are vocal-capable (about half carry non-empty lyrics fields), whereas the larger MusicPrefs and AIME pools are predominantly instrumental in their outputs (MusicPrefs $100\%$ from instrumental-only generators, AIME ${\sim}69\%$). The score is identically defined on vocal inputs, and the released checkpoint can be applied to vocal generations directly. We leave a vocal-mode best-of-$N$ study (with a vocal-capable backbone and a vocal-music reference set) to future work. Population-level external validation on real singing-voice MOS data (SingMOS-Pro, SVCC~$2025$) is reported in \S\ref{par:external_vocal_validation}.

\paragraph{Why our best-of-$N$ is monotone where prior work saturates.}\label{app:bon_monotone}
CMI-RewardBench~\cite{ma2026cmirewardbench} reports best-of-$N$ saturation on SAO-small with non-monotone Top-$k$, while our four-backbone sweep ($1.1$--$3.3$\,B) shows strict Top-$1$ monotonicity in Reward. The two findings are not in tension: we evaluate the same Reward used to select (vs.\ CMI-RewardBench's cross-model transfer to MuQ-MuLan / Audiobox / SongEval), report Top-$1$ (vs.\ Top-$k$ averages), and sweep larger backbones with more spread for the selector to exploit. The two settings are complementary: ours optimizes the in-distribution selection signal, theirs stresses cross-metric transfer.

\paragraph{Mode-collapse diagnostic for the $N{=}32$ MAD rise.}\label{app:mode1_diversity}
The $N{=}32$ MAD rise on AudioLDM2-music and ACE-Step Turbo Continuous (relative to their respective $N{=}8$ and $N{=}16$ minima) could reflect either narrowing diversity or distributional drift. Mean pairwise cosine distance among the $100$ top-$1$ MERT embeddings rises $+39\%$ on AudioLDM2-music ($0.151 \to 0.209$) and $+66\%$ on ACE-Step Turbo Continuous ($0.092 \to 0.153$) from $N{=}1$ to $N{=}32$, while the two MusicGen variants stay flat ($\pm 10\%$ of $N{=}1$). The picks spread \emph{more} at higher $N$, refuting mode collapse and indicating that the MAD rise reflects distributional drift away from SDD-$706$ on the two backbones with the largest reward headroom.

\section{Mode 3 Ablations: Multi-Round Expert Iteration}
\label{app:mode3_ablations}

\paragraph{Mode 3 protocol details.}
Fine-tuning uses the AdamW~\cite{loshchilov2019adamw} optimizer with batch size $16$. We keep an exponential moving average (EMA) snapshot of the model and use the iter-$5$K EMA weights for inference. Inference uses classifier-free guidance (CFG)~\cite{ho2022cfg} at scale $4.5$ with $25$ Euler steps, applied identically to the baseline and the post-trained checkpoint.

We probe one design knob of the Section~\ref{sec:apps_mode3} expert-iteration loop beyond the learning-rate sweep already in Table~\ref{tab:apps_mode2_3} (bottom): the number of rounds. We use the same MeanAudio FluxAudio-S starting checkpoint, the same SDD-$100$ prompts, and the same scoring protocol. The fine-tune training loss at iter $5$\,K decreases monotonically with the learning-rate sweep ($10^{-6}{:}\,0.68$, $5{\times}10^{-6}{:}\,0.48$, $10^{-5}{:}\,0.27$).

\paragraph{Multi-round expert iteration probe.}
Starting from the conservative $10^{-6}$ single-round endpoint, we run two further rounds of the full generate / score / filter / fine-tune loop at the same learning rate. The reward signal collapses round over round (Table~\ref{tab:mode3_multi_round}): mean reward drops from ${-}0.096$ (R$1$) to ${-}0.222$ (R$2$) to ${-}0.427$ (R$3$, below the R$0$ baseline ${-}0.262$), with Win shrinking from $67$ to $41$ of $100$, and MAD drifts monotonically away from SDD-$706$ (the CLAP score stays approximately flat, ${\sim}{+}0.02$ across rounds). Each round's top-decile filter draws from an already fine-tuned (narrower) backbone, and with the learning rate held fixed across rounds the fine-tune step has no mechanism to broaden the post-filter distribution. Within the configurations we tried, single-round fine-tuning consistently outperforms multi-round iteration at this learning rate (the $5{\times}10^{-6}$ single-round point identified in Section~\ref{sec:apps_mode3} as the most favorable swept learning rate remains our recommended setting), and iterating without an explicit diversity preserver (e.g., a KL anchor to the R$0$ backbone) hurts.

\begin{table}[h]
\centering
\caption{Mode~3 multi-round expert iteration at learning rate $10^{-6}$, on the public MeanAudio FluxAudio-S checkpoint. Each round is the same generate / score / filter / fine-tune loop ($900$ candidates, top-decile filter, $5$\,K iterations) initialized from the previous round's endpoint. \emph{Top-$90$ filter mean} is the mean reward of the $90$ expert samples retained by that round's filter. Parenthesized values are the change from R$0$, computed before rounding. R$3$'s mean reward sits below the R$0$ baseline ${-}0.262$, confirming the reward signal collapses across rounds.}
\label{tab:mode3_multi_round}
\footnotesize
\begin{tabular}{lccccc}
\toprule
Round & Reward$\uparrow$ & MAD$\downarrow$ & CLAP score$\uparrow$ & Win & Top-$90$ filter mean \\
\midrule
R$0$ baseline & $-0.262$ & $1.758$ & $0.0921$ & -- & -- \\
R$1$ & $-0.096$ (${+}0.166$) & $2.051$ (${+}0.293$) & $0.1109$ (${+}0.019$) & $67/100$ & ${+}0.728$ \\
R$2$ & $-0.222$ (${+}0.040$) & $2.594$ (${+}0.836$) & $0.1138$ (${+}0.022$) & $48/100$ & ${+}0.677$ \\
R$3$ & $-0.427$ (${-}0.165$) & $2.736$ (${+}0.978$) & $0.1109$ (${+}0.019$) & $41/100$ & ${+}0.577$ \\
\bottomrule
\end{tabular}
\end{table}

\section{Released Artifacts and License Interplay}
\label{app:release_detail}

\paragraph{Released artifacts.}
\begin{itemize}[leftmargin=*,itemsep=2pt]
\item \textbf{TuneJury checkpoint.} The released $2048$-d CLAP$+$MERT variant (Section~\ref{sec:tunejury}) backs all numbers in Sections~\ref{sec:evaluation}--\ref{sec:applications} and the appendix. CC-BY-NC~$4.0$, tracking the MERT-v$1$-$330$M~\cite{li2024mert} upstream license.
\item \textbf{Auxiliary checkpoints.} Leave-one-dataset-out CLAP$+$MERT variants (leave-MA, MP, AIME, SE), double-leave-out variants (leave-(SE$+$MA), leave-(MP$+$MA)), and a MuQ-MuLan-large encoder-swap variant (leave-MA $3$-dataset mix). All CC-BY-NC~$4.0$, design-space ablations in Appendix~\ref{app:cmi_external_detail}.
\item \textbf{Codebase.} Feature-extraction pipelines (LAION-CLAP and MERT, with MuQ-MuLan as an alternative), the pairwise-logistic training loop, the held-out evaluation harness, and runnable demos for Mode~1 best-of-$N$ (four frozen backbones, Section~\ref{sec:apps_bon}), Mode~2 DITTO (SAO-small and TangoFlux, Section~\ref{sec:apps_mode2}), and Mode~3 expert iteration on FluxAudio-S (Section~\ref{sec:apps_mode3}).
\item \textbf{Pre-computed reward scores} for a ${\sim}219$\,K-track open-license pool: MTG-Jamendo~\cite{bogdanov2019mtg} (${\sim}55.7$\,K), FMA-Large~\cite{defferrard2017fma} (${\sim}106$\,K), MagnaTagATune (MTAT)~\cite{law2009mtat} (${\sim}26$\,K), OpenMIC~\cite{humphrey2018openmic} ($20$\,K), MidiCaps~\cite{melechovsky2024midicaps} ($5$\,K, FluidSynth-rendered), MusicCaps~\cite{agostinelli2023musiclm} (${\sim}5.4$\,K), and the Song Describer Dataset~\cite{manco2023sdd} ($706$). Each track has one track-level reward: the CLAP branch encodes the centre $10$\,s window, the MERT branch averages the full track, and the text branch receives a $512$-d zero vector (the empty-prompt protocol). Clip-level and vocal-removed variants are planned.
\end{itemize}

\paragraph{Text-branch input for release scoring.}
The seven release datasets carry heterogeneous text formats (multi-label tags, artist/title metadata, LLM captions, human descriptions). To keep the release uniform, we feed an \emph{empty string} to the text branch under the $512$-d zero-vector protocol of Section~\ref{sec:tunejury}, releasing one \texttt{(audio, empty-prompt)} reward column that downstream users can re-score with their own prompts.

\begin{figure}[h]
\centering
\includegraphics[width=\linewidth]{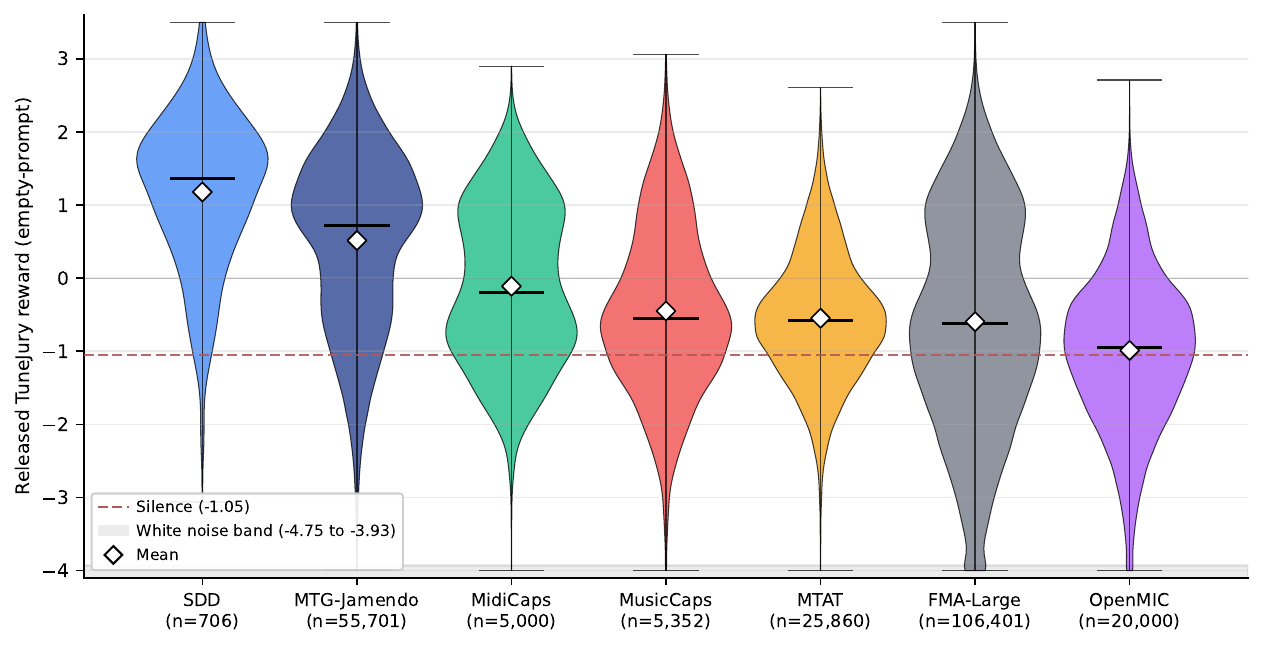}
\caption{TuneJury reward distribution across the seven release datasets (extreme tails clipped at $[-4, +3.5]$). All sources are human-music collections except MidiCaps (symbolic MIDI rendered via FluidSynth FluidR3\_GM). Black bars are medians, white diamonds are means, the dotted gray line is the silence baseline, and the shaded gray band is the white-noise baseline range. Sample counts $n$ below each violin.}
\label{fig:release_distribution}
\end{figure}

\paragraph{Reward distribution per dataset.}
Figure~\ref{fig:release_distribution} shows the per-dataset reward distribution across the released collection, with numerical complement in Table~\ref{tab:release_distribution}.

\begin{table}[h]
\centering
\caption{Per-dataset reward statistics for the released
collection (numerical complement to Figure~\ref{fig:release_distribution}).}
\label{tab:release_distribution}
\footnotesize
\begin{tabular}{lcccccc}
\toprule
Dataset & $N$ & Mean & Std & $P_{10}$ & Median & $P_{90}$ \\
\midrule
Song Describer Dataset~\cite{manco2023sdd} & $706$ & $+1.179$ & $1.038$ & $-0.305$ & $+1.364$ & $+2.341$ \\
MTG-Jamendo~\cite{bogdanov2019mtg} & $55{,}701$ & $+0.515$ & $1.205$ & $-1.168$ & $+0.718$ & $+1.934$ \\
MidiCaps~\cite{melechovsky2024midicaps} & $5{,}000$ & $-0.110$ & $1.080$ & $-1.506$ & $-0.193$ & $+1.340$ \\
MusicCaps~\cite{agostinelli2023musiclm} & $5{,}352$ & $-0.449$ & $1.122$ & $-1.841$ & $-0.547$ & $+1.139$ \\
MTAT~\cite{law2009mtat} & $25{,}860$ & $-0.548$ & $0.866$ & $-1.619$ & $-0.583$ & $+0.626$ \\
FMA-Large~\cite{defferrard2017fma} & $106{,}401$ & $-0.596$ & $1.477$ & $-2.538$ & $-0.623$ & $+1.347$ \\
OpenMIC~\cite{humphrey2018openmic} & $20{,}000$ & $-0.988$ & $1.045$ & $-2.336$ & $-0.948$ & $+0.325$ \\
\bottomrule
\end{tabular}
\end{table}

SDD sits highest (curated MTG-Jamendo provenance, mean $+1.179$, median $+1.364$), MTG-Jamendo second (professional catalog), MidiCaps third under uniform FluidSynth synthesis. MTAT, MusicCaps, FMA-Large, and OpenMIC all center below zero, with FMA-Large the broadest ($\mathrm{std} = 1.477$, $P_{10} = -2.538$) and OpenMIC the lowest-centered (mean $-0.988$, $P_{10} = -2.336$). No dataset's $90$th percentile sits below the silence / white-noise baseline (Appendix~\ref{app:sanity}), so a single global threshold $\tau$ separating ``music'' from ``broken'' inputs is plausible across these seven collections (we do not validate $\tau$ on held-out data).

\paragraph{Within-dataset reward drivers.}
The seven distributions hide reproducible within-dataset structure. On MTG-Jamendo, genre / mood / instrument tags span roughly $2.0$~reward units (happy / folk / jazz at the top vs.\ industrial / experimental at the bottom). On MidiCaps, tempo and duration are flat ($|r| < 0.03$), but major-mode tracks score reliably above minor-mode (Welch $t = 7.43$, $p < 10^{-3}$). The largest within-dataset effect is on OpenMIC: guitar / piano / ukulele / violin / mandolin clips score in $[-0.56,\,-0.43]$, while voice / drums / synthesizer clips score in $[-1.53,\,-1.10]$, a gap of ${\sim}0.8$ reward units consistent with an \emph{instrumentation prior}. Practitioners filtering OpenMIC-style heterogeneous collections should condition on instrument label.

\paragraph{Full-track vs.\ clip-level scoring.}
The release column scores each track end-to-end. A sliding-window probe ($10$\,s windows, $5$\,s hop) on $8$ MTG-Jamendo tracks ($\ge 60$\,s) shows average within-track spread of $2.28$ reward units, with the worst $10$\,s window across the $8$ tracks averaging $-1.53$ vs.\ full-track $+0.03$. The released column is therefore reasonable for cross-dataset distributional statistics but smooths over localized artifacts, so practitioners filtering for uniformly good tracks should layer a sliding-window rescore.

\paragraph{Soundfont sensitivity for the MidiCaps stream.}
Because MidiCaps is symbolic, its reward column reflects both the score and the synthesizer. Re-rendering the first $300$ MidiCaps tracks with a low-fidelity General~MIDI bank (\href{https://github.com/craffel/pretty-midi/blob/main/pretty_midi/TimGM6mb.sf2}{TimGM6mb}, $5.7$\,MB) instead of the default \href{https://packages.debian.org/sid/fluid-soundfont-gm}{FluidR3\_GM} ($142$\,MB) under the same FluidSynth front-end gives modestly higher means (paired $t = 1.73$, $p \approx 0.085$) and noisy track-level rankings (cross-soundfont Spearman $+0.69$, only $33\%$ of FluidR3\_GM top-$10\%$ tracks remain in TimGM6mb top-$10\%$). Distribution-level comparisons are safe, while track-level rankings should be treated as $(\text{score},\,\text{renderer})$ joint quantities. The renderer is documented in the release metadata.

\paragraph{License interplay.}
TuneJury is released under CC-BY-NC~$4.0$, tracking the strictest upstream constraint (MERT-v$1$-$330$M~\cite{li2024mert} weights). Training-source licenses: Music Arena~\cite{kim2025musicarena} (CC-BY~$4.0$), MusicPrefs~\cite{huang2025musicprefs} (released open-source by its authors), AIME~\cite{grotschla2025aime} (CC-BY~$4.0$), SongEval~\cite{yao2025songeval} (CC-BY-NC-SA~$4.0$). A commercial-friendly Apache~$2.0$ variant trained only on LAION-CLAP-Music audio embeddings~\cite{wu2023clap} (Row~A$1$ in Appendix~\ref{app:feature_modality}, $0.705$ overall, tied with the seed-matched A$7$ retrain) is also released. Per-backbone licenses for Modes~1--3 are documented in the release repository.

\paragraph{Use cases.}
\emph{(i)~Best-of-$N$ selection} (Section~\ref{sec:apps_bon}); \emph{(ii)~DITTO-style latent optimization} with full-sampler or late-stage backprop, base weights frozen (Section~\ref{sec:apps_mode2}); \emph{(iii)~Reward-ranked supervised fine-tuning (SFT) post-training} (expert iteration / ReST; Section~\ref{sec:apps_mode3}); \emph{(iv)~Quality-aware dataset filtering} via $\text{TuneJury} > \tau$ before generative training (Appendix~\ref{app:sanity}); \emph{(v)~Held-out evaluation} alongside FAD and the CLAP score on small prompt sets where distribution-level metrics miss the instance-level signal.

\section{Reproducibility Notes}
\label{app:reproducibility}

\paragraph{Training hyperparameters.}
The AdamW~\cite{loshchilov2019adamw} optimizer (learning rate $10^{-4}$, weight decay $10^{-3}$, batch size $32$), $4$-hidden-layer MLP head with widths $[1024, 512, 256, 128]$, BatchNorm~\cite{ioffe2015batchnorm} and ReLU between layers, dropout~\cite{srivastava2014dropout} $0.5$ on every hidden layer, pairwise logistic loss~\cite{burges2005ranknet} on the score difference. Up to $1{,}000$ epochs with early stopping on validation loss (patience $30$). Typical convergence is under $200$ epochs.

\paragraph{Random seed and runtime.}
TuneJury training uses seed $42$ (\texttt{torch}, \texttt{numpy}, \texttt{random}, \texttt{torch.backends.cudnn.deterministic=True}). A full training run completes in roughly $10$ minutes on a single NVIDIA RTX A$5000$.

\paragraph{Encoder feature extraction.}
LAION-CLAP-Music: $48$\,kHz mono input fed through the music checkpoint, with the $512$-d audio and $512$-d text projection outputs concatenated. MERT-v$1$-$330$M: $24$\,kHz mono input, last hidden state averaged over the time dimension ($1024$-d). All features are pre-extracted to disk before training, so per-step compute is the MLP head only. For SongEval training pairs and any inference-time empty-prompt call, the text branch receives a $512$-d zero vector in place of the CLAP text embedding (Section~\ref{sec:tunejury}; the released \texttt{tunejury.Scorer.score(audio, prompt="")} entry point handles this routing internally). All reported numbers use torch $2.4.0$, torchaudio $2.4.0$, and transformers $4.44.0$. Newer stacks shift the frozen-encoder outputs slightly (under torch $2.7$, CLAP cosines move by ${\sim}0.02$ and mean rewards by ${\sim}0.05$) while preserving signs, orderings, and win counts.

\paragraph{Bench-clean Music Arena UUIDs.}
The $1{,}340$ CMI-RewardBench MA test \texttt{battle\_uuid}s are sourced from the CMI-RewardBench release~\cite{ma2026cmirewardbench}. We remove the full set from our entire MA pool (train, validation, held-out test) before constructing TuneJury training splits.

\paragraph{Mode-specific configurations.}
Mode~1 backbones use library-default sampling with only the noise seed varying across candidates. Mode~2 runs full $8$-step sampler backprop on SAO-small and TangoFlux (Section~\ref{sec:apps_mode2}). AudioLDM2-music is omitted because its $50$-step UNet backprop is memory-prohibitive on our hardware. Mode~3 uses CFG $4.5$, $25$ Euler steps, no post-processing (Section~\ref{sec:apps_mode3}). All MAD values use \texttt{mauve.compute\_mauve} (\texttt{num\_buckets=\textquotesingle auto\textquotesingle}, \texttt{seed=42}) on $1024$-d MERT time-mean embeddings against SDD-$706$. The released code documents per-experiment sampling configs and the SDD-$100$ subset prompt list.

\paragraph{SAO-small Mode 2 sample-size and snapshot note.}\label{app:sao_caveat}
SAO-small Mode~2 numbers in Table~\ref{tab:apps_mode2_3} are computed at $n{=}30$ on a \href{https://github.com/Stability-AI/stable-audio-tools}{\texttt{stable-audio-tools}} $0.0.18$ snapshot (May~$2026$). The current release exceeds $24$\,GB working memory at $n{=}100$ ($10$\,s $/$ $44.1$\,kHz with default CFG~$6$), and step-level gradient checkpointing / bf16 / sequential CFG each move the OOM site without bringing peak below $24$\,GB. The reproducer pipeline at this snapshot (determinism settings: cuDNN deterministic kernels, math-only SDPA, \texttt{use\_deterministic\_algorithms}) yields the Table~\ref{tab:apps_mode2_3} SAO-small row up to ${\sim}0.05$ run-to-run reward-lift variance on the SAO-small autograd path. Absolute baseline / post-DITTO values and the sign of the MAD change depend on the sampler snapshot, so different \texttt{stable-audio-tools} versions may shift these cells. TangoFlux Mode~2 (no CFG by design) and the single-pass Mode~1 / Mode~3 paths are unaffected.

\end{document}